\documentclass[nohyper]{JHEP3}
\usepackage{latexsym}
\usepackage{amssymb}
\usepackage{graphicx}
\usepackage{multibox}
\usepackage{amsmath,amsfonts}

\def\*{\ast}

\def\ve{\varepsilon}

\def\be{\begin{equation}}
\def\ee{\end{equation}}
\def\bqn{\begin{eqnarray}}
\def\eqn{\end{eqnarray}}

\def\theequation{\thesection.\arabic{equation}}

\newsavebox{\ver}
\newsavebox{\verp}

\newsavebox{\gorp}
\newsavebox{\toch}

\newcommand{\bee}{\begin{eqnarray}}
\newcommand{\eee}{\end{eqnarray}}

\date{}

\title{The complete $AdS_4\times CP^3$   superspace for  the type IIA
superstring and D-branes}
\author{Jaume Gomis\footnote{\tt jgomis@perimeterinstitute.ca}\ , Dmitri~Sorokin\footnote{\tt dmitri.sorokin@pd.infn.it, linus.wulff@pd.infn.it}$^\dagger$ and Linus Wulff$^{\hskip+1pt\dagger}$
~\\
~\\
~\\
{\it $^*$Perimeter Institute for Theoretical Physics}
~\\
{\it Waterloo, Ontario N2L 2Y5, Canada}
~\\
~\\
{\it $^\dagger$Istituto Nazionale di Fisica Nucleare, Sezione di Padova,}
~\\
{\it via F. Marzolo 8, 35131 Padova, Italia}}

\abstract
{ We lift the bosonic $AdS_4\times CP^3$ solution of type IIA
supergravity preserving 24 supersymmetries to a $D=10$ superspace
which has 32 Grassmann--odd directions. The type IIA superspace is
obtained from $D=11$ via dimensional reduction of the coset
superspace $OSp(8|4)/SO(7)\times SO(1,3)$ by realizing the latter as
a Hopf fibration over the former. This construction generalizes to
superspace the Hopf fibration of $S^7$ as a $U(1)$ bundle over
$CP^3$, and is suitable for writing the explicit form of
Green--Schwarz--type actions encoding the dynamics of the type IIA
string and branes in the $AdS_4\times CP^3$ superbackground. We show
that the $OSp(6|4)/U(3)\times SO(1,3)$ supercoset string action
describes only a subsector of the complete Green--Schwarz
superstring. Thus, even though the superstring equations of motion
in the $OSp(6|4)/U(3)\times SO(1,3)$ subsector are classically
integrable, the fact that the full $AdS_4\times CP^3$ superspace is
not a supercoset requires the use of more general methods to
determine whether the superstring in the complete $AdS_4\times CP^3$
superbackground is classically integrable.}

\begin{document}
\section{Introduction}
Recent progress in understanding the $AdS_4/CFT_3$ correspondence
has been triggered by the construction of
Bagger-Lambert-Gustavsson-type models based on tri--algebras
\cite{Bagger:2006sk,Gustavsson:2007vu,Gomis:2008uv}
and by the  model due to Aharony, Bergman, Jafferis and Maldacena
(ABJM) \cite{Aharony:2008ug}.\footnote{The ABJM Lagrangian is a
special case of the ${\cal N}=4$ superconformal Chern-Simons
theories written down in \cite{Gaiotto:2008sd}.} These new models
--- based on 3--dimensional ${\cal N}$--extended superconformal
Chern--Simons gauge theories coupled to scalar supermultiplets ---
have been conjectured to provide an effective low energy description
of multiple coincident M2--branes in  M--theory, with the ABJM
theory at level $k$ describing the physics of multiple M2--branes on
an $R^8/Z_k$ orbifold \cite{Aharony:2008ug}. These novel three
dimensional theories provide us with new tools for studying the
$AdS_4/CFT_3$ duality from the boundary field theory point of view,
and may shed new light on the landscape of $AdS_4$ vacua in string
theory.

The ${\cal N}=6$ Chern-Simons theory with gauge group $U(N)_k\times
U(N)_{-k}$ constructed in \cite{Aharony:2008ug} describes M-theory
on $AdS_4 \times S^7/Z_k$. There is a region in the parameter space
of the ABJM theory\footnote{Corresponding to $N^2>>\lambda^{5/2}$,
where $\lambda$ is the 't Hooft coupling of the ABJM theory.} where
the bulk description is given in terms of perturbative type IIA
string theory on the $AdS_4 \times CP^3$ background, which preserves
24 out of 32 supersymmetries. Therefore, in order to study this new
type of holographic correspondence using the bulk description,  one
needs an explicit form of the superstring action on the type IIA
superspace whose bosonic body is $AdS_4
\times CP^3$. Likewise, writing down the action of D-branes on the
$AdS_4
\times CP^3$ superbackground is useful, as D-branes in   $AdS_4
\times CP^3$ play an important role in the duality, since they
describe various local and non-local operators in the dual gauge
theory
\cite{Aharony:2008ug,Drukker:2008zx,Drukker:2008jm,Nishioka:2008ib,Berenstein:2008dc,Kluson:2008wn}.
Of course, the Green--Schwarz--type form of the superstring action
and superbrane actions in generic superbackgrounds are well known
\cite{Green:1983wt,Grisaru:1985fv,Bergshoeff:1987cm,Cederwall:1996pv,Aganagic:1996pe,Bergshoeff:1996tu,Bandos:1997rq,Bandos:1997ui}.
The challenge  is to obtain the explicit form of the superstring and
superbrane actions\footnote{The superstring action to   quadratic
order in the fermionic coordinates is known in an arbitrary
superbackground \cite{Cvetic:1999zs}.} for the various $AdS_4/CFT_3$
superbackgrounds, by finding the explicit dependence of the
supervielbeins, NS--NS and RR superfields on the 32 fermionic
coordinates of the type IIA superbackground of interest.

Analogous demand for explicit actions for the superstring  and
branes arose in the early studies of the $AdS_5/CFT_4$ and
$AdS_4/CFT_3$ correspondence. In the maximally supersymmetric
$AdS_5\times S^5$  superbackground, the supergeometry is  described
by the coset superspace $SU(2,2|4)/SO(5)\times SO(1,4)$, and the
explicit form of the action for the type IIB superstring was found
in
\cite{Metsaev:1998it,Kallosh:1998zx} while  the D3--brane action was constructed in \cite{Metsaev:1998hf}.
Analogous actions were derived for the M2--brane \cite{deWit:1998tk}
and the M5--brane
\cite{Claus:1998fh,Pasti:1998tc} in the $AdS_4\times S^7$ and $AdS_7\times S^4$
superbackgrounds respectively, which are  described by the
supercosets $OSp(8|4)/SO(7)\times SO(1,3)$ and
$OSp(6,2|4)/SO(4)\times SO(1,6)$.

The construction of the superstring and brane actions in the
$AdS_4\times CP^3$ background is significantly more complicated, as
the background preserves only 24 out of the 32 supersymmetries of type
IIA supergravity. A  coset superspace whose isometries are those of
the $AdS_4\times CP^3$ vacuum is $OSp(6|4)/U(3)\times SO(1,3)$. Its
bosonic body is the desired $AdS_4\times CP^3$ geometry  and its
Grassmann--odd subspace is 24--dimensional.  Therefore,
$OSp(6|4)/U(3)\times SO(1,3)$ is  a particular solution of the type
IIA supergravity constraints which can be regarded as a submanifold
in  the general $AdS_4\times CP^3$ IIA superspace, whose
Grassmann--odd sector is 32--dimensional.

A sigma--model action for the superstring propagating in the
$OSp(6|4)/U(3)\times SO(1,3)$ submanifold of the complete type IIA
superspace  was constructed and analyzed in
\cite{Arutyunov:2008if,Stefanski:2008ik,Fre:2008qc,Bonelli:2008us,D'Auria:2008cw}.
This action can be regarded as the Green--Schwarz action for the
superstring in an $AdS_4\times CP^3$ superspace with 32 fermionic
directions in which the 16--parameter kappa--symmetry has been
partially fixed in order to eliminate the 8 fermionic coordinates of
the string corresponding to the 8 broken supersymmetries. With this
interpretation, only 24 fermionic modes on the string worldsheet
remain and these are described by the sigma--model based on the
$OSp(6|4)/U(3)\times SO(1,3)$ supercoset. This fixing of
kappa--symmetry restricts the motion of the string to a submanifold
of bosonic dimension 10 and fermionic dimension 24 in the total type
IIA superspace. As already noted in \cite{Arutyunov:2008if}, the
$OSp(6|4)/U(3)\times SO(1,3)$ sigma--model action does not describe
all possible motions of the string in the $AdS_4\times CP^3$
superspace. In particular, if the string moves entirely in $AdS_4$,
the number of kappa--symmetries of this sigma--model gets increased
from 8 to 12. This indicates that this dynamical sector of the
theory cannot be attained from the gauge choice for fixing
kappa--symmetry of the Green--Schwarz string action that yields the
coset superspace. In this sector of the theory,  four of the modes
associated with the eight broken supersymmetries are dynamical
fermionic degrees of freedom of the superstring. The reason behind
this is that when the string moves entirely in $AdS_4$,  its
kappa--symmetry projector commutes with the projector which singles
out the 8 broken supersymmetries,  and therefore it cannot eliminate
all the corresponding fermionic modes but only half of them.

Therefore, the study of the general classical and quantum motion of
the superstring in $AdS_4\times CP^3$ cannot be achieved using the
$OSp(6|4)/U(3)\times SO(1,3)$ supercoset. We need to find an action
that includes the extra dynamical fermionic modes. On general
grounds, this is given by the Green--Schwarz superstring action in
the  $AdS_4\times CP^3$ superspace with 32 Grassmann--odd
coordinates coupled to a corresponding NS--NS 2--form superfield
depending on 32 $\theta$s. In this paper we present this action.

Likewise, a D2--brane which is embedded purely in an $AdS_4$
subspace\footnote{An example of this situation is the D2--brane with
$AdS_2\times S^1\subset AdS_4$ worldvolume
\cite{Drukker:2008jm}, which corresponds to a disorder loop operator
in the ABJM theory, and another example is the D2--brane at the
Minkowski boundary of $AdS_4$.} of $AdS_4\times CP^3$ cannot be
described by the D2--brane action based on the $OSp(6|4)/U(3)\times
SO(1,3)$ supercoset,  since the embedding  is incompatible with the
kappa--symmetry gauge fixing\footnote{The discussion of the problem
of fixing $\kappa$--symmetry in the D0-- and D2--brane actions in
$AdS_4\times CP^3$ superspaces has been done in collaboration with
P. Fr\'e and P.A. Grassi.\label{fg}} of the corresponding
Green--Schwarz--type D2--brane action
\cite{Cederwall:1996pv,Aganagic:1996pe,Bergshoeff:1996tu}. Other
examples of this situation are D2- and D4--branes partially moving
in $AdS_4$ and wrapping the 2--cycle in $CP^3$ associated with the
$CP^3$ K\"ahler form $J$. Thus, to describe a  general D--brane
configuration  in $AdS_4\times CP^3$ one needs once again an
explicit form of its action in the $AdS_4\times CP^3$ superspace
with 32 Grassmann--odd coordinates coupled to the corresponding
NS--NS and RR superfields depending on 32 $\theta$s.

The main  result of this paper is the explicit  construction of  the
complete $AdS_4\times CP^3$ superspace including all of the 32
Grassmann--odd coordinates. Unlike for most of the supergeometries
studied previously in the literature, this type IIA $AdS_4\times
CP^3$ superspace is not a coset superspace, but we can nevertheless
completely characterize its supergeometry. Having determined the
supervielbeins of this superspace and the corresponding NS--NS and
RR gauge superfields, we explicitly write down the general
Green-Schwarz-type actions for the type IIA superstring and
D--branes in $AdS_4\times CP^3$. We analyze the classical equations
of motion of the superstring in different submanifolds of the
$AdS_4\times CP^3$ superspace. On the submanifold described by the
$OSp(6|4)/U(3)\times SO(1,3)$ coset superspace, the classical
superstring equations of motion are integrable
\cite{Arutyunov:2008if,Stefanski:2008ik}, generalizing the corresponding
result found by Bena, Polchinski and
Roiban for the type IIB superstring propagating on the $AdS_5\times S^5$ supercoset
\cite{Bena:2003wd}. However,
we find that there is a submanifold  in the $AdS_4\times CP^3$
superspace   that is described by a ``twisted" $OSp(2|4)/SO(2)\times
SO(1,3)$ superspace, which is not a supercoset, and the ingredients
used to prove integrability found in
\cite{Bena:2003wd} do not directly apply to this sector of the theory.
Therefore, it remains an important open problem to determine whether
the complete set of classical equations of motion of the
Green-Schwarz superstring propagating on the $AdS_4\times CP^3$
superspace is classically integrable. The fact that the $AdS_4\times
CP^3$ superspace with 32 fermionic directions is not a supercoset
requires more general techniques to prove classical integrability.

The explicit form of the supervielbeins and superconnections
describing the $AdS_4\times CP^3$ superspace are obtained by
performing the Kaluza--Klein reduction of the supergeometry of the
supercoset $OSp(8|4)/SO(7)\times SO(1,3)$, which is a solution of
the D=11 superfield supergravity constraints corresponding to the
maximally supersymmetric $AdS_4\times S^7$ vacuum of eleven
dimensional supergravity. It is well known since the first intensive
studies  of flux compactifications of D=10 and D=11 supergravities
that    type  IIA supergravity vacua\footnote{Let us here make the
historical remark that the compactified vacuum solutions of type IIA
supergravity corresponding to a direct product of $AdS_4$ and a
compact manifold $M^6$
\cite{Watamura:1983ht,Volkov:1984yw,Campbell:1984zc,Watamura:1983hj} were obtained by a combination
of two mechanisms of spontaneous (flux) compactification proposed in
1980. One of the mechanisms was due to Freund and Rubin
\cite{Freund:1980xh} in which the compactification of a D--dimensional theory
into an $AdS_n\times M^{D-n}$ manifold takes place as a result of
the interaction of gravity with a closed $n$--form or $(D-n)$--form
field strength of an antisymmetric gauge field. Another mechanism
was proposed by Volkov and Tkach
\cite{Volkov:1980kq}. Volkov and Tkach showed that in an interacting theory of gravity
with Yang--Mills fields the compactification of extra dimensions may
take place into coset spaces when components of the Yang--Mills
fields take the same values as some components of the spin
connection of the compactified manifold. The field strengths of the
vacuum configurations of the Yang--Mills fields are (using the
modern terminology)  topologically nontrivial fluxes supported by
compact subspaces.} can be lifted to corresponding bosonic solutions
of D=11 supergravity by constructing $U(1)$ fibrations over the ten
dimensional manifold  characterizing the type IIA supergravity
solutions
\cite{Giani:1984wc,Nilsson:1984bj,Sorokin:1985ap}.
For example,  the 7--sphere is a $U(1)$ Hopf fibration over $CP^3$,
and therefore the  $AdS_4\times CP^3$ solution of the bosonic
equations of type IIA supergravity
\cite{Watamura:1983hj} is directly related
to the Freund--Rubin $AdS_4\times S^7$ solution of the bosonic
$D=11$ supergravity equations of motion by  reducing along the
$U(1)$--fiber direction of the $S^7$
\cite{Nilsson:1984bj,Sorokin:1985ap}. For recent generalizations of
these old results to the description of new compactified type IIA
vacua see \emph{e.g.}
\cite{Aldazabal:2007sn,Tomasiello:2007eq,Koerber:2008rx,Martelli:2008rt}.

Extending  the Kaluza-Klein reduction to superspace is much more
subtle. When the Hopf fibration of $AdS_4\times S^7$ is lifted to
$D=11$ superspace, such that $AdS_4\times S^7$ becomes the bosonic
subspace of the $OSp(8|4)/SO(7)\times SO(1,3)$ supercoset, the
supervielbeins of the supercoset do not come in  a  form suitable
for performing the dimensional reduction of the D=11 superspace down
to the type IIA D=10 superspace  (see
\cite{Duff:1987bx} for the general prescription for performing such a
superspace reduction and
\cite{Howe:2004ib} for more details). As we shall show, to get the
$OSp(8|4)/SO(7)\times SO(1,3)$ supervielbeins in the
Kaluza--Klein--like form one should perform a ``twist" of their
components along the $AdS_4$ and the $U(1)$--fiber directions, or in
other words   perform a local Lorentz rotation in the 5--dimensional
subspace tangent to $AdS_4$ and the $U(1)$--fiber direction along
$S^7$. We should stress that such a transformation is not part of
the isometry of the $AdS_4\times S^7$ solution and should be
regarded as   an appropriate choice of a different supervielbein
basis of $OSp(8|4)/SO(7)\times SO(1,3)$ which has the Kaluza--Klein
form compatible with the Hopf fibration. Note that by orbifolding
the $OSp(8|4)/SO(7)\times SO(1,3)$ supercoset by $Z_k\subset U(1)$,
where $U(1)$ is the commutant of $SU(4)$ in $SO(8)$, one gets the
supergeometry corresponding to the superspace with an $AdS_4\times
S^7/Z_k$ bosonic subspace, a background of eleven dimensional
supergravity which preserves 24 supersymmetries (for $k>2$) and is
the near horizon geometry of N M2--branes probing the $C^4/Z_k$
singularity.

Having obtained the complete supergeometry with 32 fermionic
directions   describing the $AdS_4\times CP^3$ solution of type IIA
supergravity,  one can then use it to write down the
Green--Schwarz--type actions for the type IIA superstring and
D--branes (or the pure spinor action for the superstring) depending
on all 32 fermions. This gives the complete and consistent
description of these objects in the type IIA $AdS_4\times CP^3$
superbackground. The complete form of the Green--Schwarz  action
provides a systematic framework in which to study the $AdS_4/CFT3$
correspondence and other problems.

The plan of the rest of the paper is as follows. In Section
\ref{superadscp3}, for the reader's convenience, we summarize
our results and write down the explicit supergeometry for the type
IIA $AdS_4\times CP^3$  background. The details of our computations
appear in the rest of the paper. In Section \ref{actions} we write
down the actions for the superstring and D-branes in this
superbackground. We also analyze the motion of the string in
submanifolds of the $AdS_4\times CP^3$  superspace and note that the
string equations of motion in a certain subspace are integrable
\cite{Arutyunov:2008if,Stefanski:2008ik}. We find, however, that there is a
submanifold in superspace for which the criteria found to prove
integrability in \cite{Bena:2003wd} are not satisfied. So whether
the Green-Schwarz superstring in $AdS_4\times CP^3$ is integrable
remains to be proven. In Section \ref{s7fibration} we describe a
coset space realization of $S^7$ as a $U(1)$ bundle over $CP^3$. In
Section
\ref{11sgfibration} we lift the  Hopf fibration description of the
$S^7$ to $D=11$ superspace and show that the associated
supervielbeins and superconnections can be brought to the
Kaluza-Klein form by performing a particular local Lorentz
transformation, which allows us to read off the supergeometry for
the type IIA $AdS_4\times CP^3$ background. The main notation,
conventions and some computations are  presented in the Appendices
A-C.

\section{$AdS_4 \times CP^3$ superspace with 32 Grassmann--odd directions}\label{superadscp3}

In this Section we summarize our main result, namely, the
construction of the superspace which has 32 Grassmann--odd
directions, contains $AdS_4
\times CP^3$ as its bosonic part and solves the type IIA
supergravity constraints
\cite{Carr:1986tk,Cederwall:1996pv,Bergshoeff:1996tu,Howe:2004ib}. The derivation of this
result is given in Sections
\ref{s7fibration}--\ref{11sgfibration}.

The type IIA superspace of interest is parametrized by 10 bosonic
coordinates $X^M=(x^m,\,y^{m'})$, where $x^m$ $(m=0,1,2,3)$ and
$y^{m'}$ $(m'=1,\cdots 6)$ parametrize $AdS_4$ and $CP^3$
respectively, and by 32-fermionic coordinates
$\theta^{\underline\mu}=(\theta^{\mu\mu'})$, which  combine  into
the supercoordinates $Z^{\cal M}=(x^m,\,y^{m'},\,\theta^{\mu\mu'})$.
The spinor indices $\mu=1,2,3,4,$ and $\mu'=1,\cdots,8$ label,
respectively, an $SO(2,3)$ and $SO(6)$ spinor representation.

The 32 fermionic coordinates $\theta^{\mu\mu'}$ split into 24
coordinates $\vartheta^{\mu m'}$, which correspond to the 24
unbroken supersymmetries of the $AdS_4 \times CP^3$ background, and
8 coordinates $\upsilon^{\mu i}$ $(i=1,2)$ corresponding to the 8
broken supersymmetries.\footnote{This splitting is carried out by
applying the projectors (\ref{p61}) and (\ref{p21}) on $\theta^{\underline\mu}$ (See Appendices A
and C for more details).}

The type IIA supervielbeins are\footnote{Our convention for the
essential torsion constraint of IIA supergravity is
$T_{\underline{\alpha\beta}}{}^A=2\Gamma_{\underline{\alpha\beta}}^A$.
This choice is related to the form of the $OSp(8|4)$ algebra
(Appendix B, eq. (\ref{QQ})) and differs from that of
\cite{Howe:2004ib} by the factor $2i$.}
\be\label{IIAsv1}
{\mathcal E}^{\mathcal A}=dZ^{\mathcal M}\,{\mathcal E}_{\mathcal
M}{}^{\mathcal A}(Z)=({\mathcal E}^{A},\,{\mathcal
E}^{\underline\alpha})\,,\qquad
\ee
where
\be\label{svA}
{\mathcal E}^{A}(Z)=({\mathcal E}^{a},\,{\mathcal E}^{a'})\,\qquad
a=0,1,2,3,\quad a'=1,\cdots,6
\ee
are the vector supervielbeins in the tangent space of $AdS_4 \times
CP^3$ and
\be\label{svalpha}
{\mathcal E}^{\underline\alpha}(Z)={\mathcal
E}^{\alpha\alpha'}=({\mathcal E}^{\alpha a'},\,{\mathcal E}^{\alpha
i})\, \qquad \alpha=1,2,3,4, \quad \alpha'=1,\cdots , 8\,,\quad
i=1,2
\ee
are the fermionic supervielbeins which split into 24 along the
unbroken supersymmetry directions and eight along the broken ones.
(The spinor indices $\alpha=1,2,3,4,$ and $\alpha'=1,\cdots,8$
label, respectively, an $SO(1,3)$ and a $U(3)$
representation.) The supervielbeins (\ref{svA}) and (\ref{svalpha})
are expressed in terms of the supervielbeins $E^{A}(x,y,\vartheta)$,
$E^{\alpha a'}(x,y,\vartheta)$ and the $U(1)$ connection
$A(x,y,\vartheta)$ of the $OSp(6|4)/U(3)\times SO(1,3)$  supercoset,
whose fermionic coordinates are $\vartheta^{\alpha a'}$, but the
former also depend on the 8 additional fermionic coordinates
$\upsilon^{\alpha i}$ as follows\footnote{These are the formulas for
the case when $k$, corresponding to the order of the $Z_k$ orbifold
of the $S^7$ and the type IIA RR two--form flux through $CP^3$,  is
set to $k=1$. The formulas for general $k$ are obtained by making
the following rescaling: $\Phi\rightarrow {1\over k}
\Phi$, $E_7^{\ a}\rightarrow {1\over k}   E_7^{\ a}$ and $e^{{2\over 3}\phi}\rightarrow {1\over
k} e^{{2\over 3}\phi}$.}
\bee
\label{sva}
{\mathcal E}^{a'}(x,y,\vartheta,\upsilon)&=&e^{{1\over
3}\phi(\upsilon)}\,\left(E^{a'}(x,y,\vartheta)-2\upsilon\,{{\sinh
m}\over m}\gamma^{a'}\gamma^5\,E(x,y,\vartheta)\right) \,,
\nonumber\\
\nonumber\\
{\mathcal E}^{a}(x,y,\vartheta,\upsilon) &=&
e^{{1\over3}\phi(\upsilon)}\,\left(E^b(x,y,\vartheta)-4\upsilon\gamma^b\,{{\sinh^2{{\mathcal
M}/ 2}}\over{\mathcal M}^2}\,D\upsilon\right)\Lambda_b{}^a(\upsilon)
\nonumber\\
&&{}
\hskip+0.5cm -e^{-{1\over3}\phi(\upsilon)}\,\left(A(x,y,\vartheta)-4i\upsilon\,\ve\gamma^5\,{{\sinh^2{{\mathcal
M}/2}}\over{\mathcal M}^2}\,D\upsilon\right) E_7{}^a(\upsilon)\,,
\nonumber\\
\\
{\mathcal E}^{\alpha i}(x,y,\vartheta,\upsilon) &=&
e^{{1\over6}\phi(\upsilon)}\,\left({{\sinh{\mathcal
M}}\over{\mathcal M}}\,D\upsilon\right)^{\beta j}\,S_{\beta
j}{}^{\alpha i}\,(\upsilon) -i\,e^{\phi(\upsilon)}{\mathcal
A}_1(x,y,\vartheta,\upsilon)\,(\gamma^{5}\ve\lambda(\upsilon))^{\alpha
i}\,,
\nonumber\\
\nonumber\\
{\mathcal E}^{\alpha a'}(x,y,\vartheta,\upsilon) &=&
e^{{1\over6}\phi(\upsilon)}\,E^{\gamma b'}(x,y,\vartheta)\,\left(
\delta_{\gamma}{}^{\beta}-8\,\left(i\gamma^5\,\upsilon\,{{\sinh^2{{m}/2}}\over{m}^2}\right)_{\gamma
i}\upsilon^{\beta i} \right)S_{\beta b'}{}^{\alpha
a'}\,(\upsilon)\,,\nonumber
\eee
where $E(x,y,\vartheta)$ in the second term of the first expression
is the spinor one--form $E^{\gamma b'}(x,y,\vartheta)$ which also
appears in the last expression of (\ref{sva}).

The type IIA RR one--form gauge superfield is
\be
\begin{aligned}
\label{RR1form}
{\mathcal A}_1(x,y,\vartheta,\upsilon) &=
e^{-{4\over3}\phi(\upsilon)}\,\left[
\left(A(x,y,\vartheta)-4i\upsilon\,\ve\gamma^5\,{{\sinh^2{{\mathcal
M}/2}}\over{\mathcal M}^2}\,D\upsilon\right)\Phi(\upsilon)
\right.\\
&\left.\hspace{40pt}+\left(E^a(x,y,\vartheta)-4\upsilon\gamma^a\,{{\sinh^2{{\mathcal
M}/2}}\over{\mathcal M}^2}\,D\upsilon\right)E_{7a}(\upsilon)
\right]\,
\end{aligned}
\ee
with the field strength $F_2=d\mathcal A_1$. The RR four-form and
NS--NS three-form superfield strengths are given by
\bee\label{f4h3}
F_4&=&d{\mathcal A}_3-{\mathcal A}_1\,H_3=-\frac{1}{4!}{\mathcal
E}^d{\mathcal E}^c{\mathcal E}^b{\mathcal
E}^a\left(6e^{-2\phi}\Phi\ve_{abcd}\right)
+\frac{1}{2}{\mathcal E}^{B}{\mathcal E}^{A}{\mathcal E}^{\underline\beta}{\mathcal E}^{\underline\alpha}e^{-\phi}(\Gamma_{AB})_{\underline{\alpha\beta}}\nonumber\\
\\
H_3&=&  dB_2=-\frac{1}{3!}{\mathcal E}^c{\mathcal E}^b{\mathcal
E}^a\left(6e^{-\phi}\ve_{abcd}E_7{}^d\right) +{\mathcal
E}^{A}{\mathcal E}^{\underline\beta}{\mathcal
E}^{\underline\alpha}\left(\Gamma_A\Gamma_{11}\right)_{\underline{\alpha\beta}}
-{\mathcal E}^{B}{\mathcal E}^{A}{\mathcal
E}^{\underline\alpha}(\Gamma_{AB}\Gamma^{11}\lambda)_{\underline\alpha}\,,\nonumber
\eee
where $\Gamma_{A}$ and $\Gamma_{11}$ are $32\times32$
gamma--matrices which in the $AdS_4\times CP^3$ background are
convenient to represent as a direct product of  $4\times4$ and
$8\times8$ gamma--matrices (see eq. (\ref{Gamma10}) of Appendix A).

 The gauge potentials of (\ref{f4h3}), which appear in the
superstring and D-brane actions, can be computed by a standard
procedure as follows:
\be\label{B2}
B_2=b_2+\int_0^1\,dt\,i_\theta H_3(x,y,t\theta)\,,\qquad \theta=(\vartheta,\upsilon)\\
\ee
\be\label{A3}
\hskip+1.9cm{\mathcal
A}_3=a_3+\int_0^1\,dt\,i_\theta\left(F_4+\mathcal{A}_1H_3\right)(x,y,t\theta)\,,
\ee
where $b_2$ and $a_3$ are the purely bosonic parts of the gauge
potentials and $i_\theta$ means the inner product with respect to
$\theta^{\mu\mu'}$. Note that $b_2$ is   pure gauge in the
$AdS_4\times CP^3$ solution. \footnote{To derive eqs. (\ref{B2}) and
(\ref{A3}) one should use the fact that the coordinate variation of
a differential superform $A(Z)=A(X,\theta)$ is $\delta A=i_{\delta
Z}dA+d(i_{\delta Z}A)$. Then, rescaling $\theta
\rightarrow t\theta$ in $A(X,\theta)$ and taking the derivative with
respect to $t$, we have $\frac{d}{dt}A(X,t\theta)=i_\theta dA
+d(i_{\theta}A)$, which upon integration over $t$ gives eqs.
(\ref{B2}) and (\ref{A3}), up to pure gauge terms.}

In eqs. (\ref{sva})--(\ref{RR1form})
\be\label{dupsilon}
D\upsilon=\left(d+iE^a(x,y,\vartheta)\gamma^5\gamma_a-\frac{1}{4}\Omega^{ab}(x,y,\vartheta)\gamma_{ab}\right)\upsilon\,,
\ee
where $E^{a'}(x,y,\vartheta)$, $E^{a}(x,y,\vartheta)$ and
$\Omega^{ab}(x,y,\vartheta)$ are, respectively, the $CP^3$ and
$AdS_4$ part of the supervielbein and the $SO(1,3)$ connection of
the  $OSp(6|4)/U(3)\times SO(1,3)$ supercoset, while  $E^{\alpha
a'}(x,y,\vartheta)$ is its spinorial supervielbein.
$A(x,y,\vartheta)$ is the $U(1)$ connection   on the
$OSp(6|4)/U(3)\times SO(1,3)$ supercoset,  which corresponds to  the
RR one--form   gauge potential for  this type IIA supergravity
solution,  while in the complete superspace it is given by (\ref{RR1form}). All these quantities are known explicitly and can be taken
in any suitable form, which one can find, e.g. in
\cite{Kallosh:1998zx,Arutyunov:2008if,Stefanski:2008ik,Fre:2008qc,D'Auria:2008cw}
or in our Appendix A, eq. (\ref{cartan24}). An appropriate choice of
the supercoset representatives may drastically simplify their
fermionic dependence (see e.g. \cite{Pasti:1998tc}).

The quantities $\Lambda_a{}^{b}(\upsilon)$ and $S_{\alpha
\alpha'}{}^{\beta\beta'}(\upsilon)$ appearing in the above equations have the form
\be
\begin{aligned}
\Lambda_a{}^{b}&= \delta_a{}^{b}-\frac{e^{-\frac{2}{3}\phi}}{e^{\frac{2}{3}\phi}+\Phi}{E_{7a}}\,E_7{}^b\,
\\
\\
S&=
\frac{e^{-\frac{1}{3}\phi}}{\sqrt2}\left(\sqrt{e^{\frac{2}{3}\phi}+\Phi}-\frac{E_7{}^a\,\Gamma_a\Gamma_{11}}{\sqrt{e^{\frac{2}{3}\phi}+\Phi}}
\,\right)
\,.
\end{aligned}
\ee
They generate the Lorentz transformation in the
$OSp(8|4)/SO(7)\times SO(1,3)$ supergeometry which brings the $D=11$
superspace into the Kaluza--Klein form required to perform its
dimensional reduction to the $D=10$ superspace (see Section 5).

The function $\phi(\upsilon)$ is the dilaton superfield of the full
type IIA superspace solution under consideration. The dilaton
superfield
depends only
on the 8 Grassmann coordinates $\upsilon^{\alpha i}$ and has the
following expression in terms of $E_7{}^a(\upsilon)$ and
$\Phi(\upsilon)$
\be\label{dilaton1}
e^{{2\over
3}\phi(\upsilon)}=\sqrt{\Phi^2+E_7{}^a\,E_7{}^b\,\eta_{ab}}\,.
\ee
The fermionic field $\lambda^{\alpha i}(\upsilon)$ describes the
non--zero components of the dilatino superfield,  which is defined
by the equation \cite{Howe:2004ib}
\be\label{dilatino1}
\lambda_{\alpha i}=-\frac{1}{3}D_{\alpha i}\,\phi(\upsilon).
\ee
Other quantities appearing in eqs. (\ref{sva})--(\ref{dilatino1}),
namely $\mathcal M$, $m$, $\Phi(\upsilon)$ and $E_7{}^a(\upsilon)$,
whose geometrical and group--theoretical meaning is explained in
Section
\ref{11sgfibration}, are explicitly    given in eqs. (\ref{U}), (\ref{phiE7}).

One can notice that a distinctive feature of the $AdS_4\times CP^3$
IIA superspace with 32 Grassmann--odd directions compared to the
coset superspace $OSp(6|4)/U(3)\times SO(1,3)$ with only 24
Grassmann--odd directions
is that in the full superspace solution the
dilaton, dilatino and the NS--NS 3--form superfield have non--zero
values, and depend on the 8 fermionic coordinates which correspond
to broken supersymmetries of the $AdS_4\times CP^3$ IIA supergravity
solution.

For brevity we do not present here the explicit form of the
superconnections of the $AdS_4\times CP^3$ superspace, since they
are not required for the construction of the superstring and brane
actions. When necessary, they can be directly recovered from the
Cartan forms of $OSp(8|4)/SO(8)\times SO(1,3)$, as explained in
Section
\ref{11sgfibration}.

\section{Actions for the type IIA superstring and D--branes in the complete $AdS_4\times CP^3$ superspace}
\label{actions}

\subsection{Type IIA Green--Schwarz superstring}
The Green--Schwarz superstring action \cite{Green:1983wt} in a
generic supergravity background is well--known \cite{Grisaru:1985fv}
and its Nambu--Goto form is

\begin{equation}\label{gs}
S=-{T}\int d^2\xi\sqrt{-\det\,({\mathcal E}_{i}{}^{A}\, {\mathcal
E}_{j}{}^{B}\,\eta_{AB})}+ T\,\int B_2 (\xi),
\end{equation}
where $T$ is the string tension, $\xi^i$ $(i=0,1)$\footnote{Since we
have exhausted a finite number of letters which are at our disposal
to define different types of indices, here we use the letters $i,j$
to denote the worldvolume indices. We believe that this will not
cause confusion with the same letters used in the previous Section
to define $SO(2)\subset SO(8)$ indices.} are the string worldsheet
coordinates, ${\mathcal E}_{i}{}^{A}=\partial_i\,Z^{\mathcal
M}(\xi)\,{\mathcal E}_{\mathcal M}{}^{A}$ is  the worldsheet
pullback of the vector supervielbeins of type IIA supergravity and
$B_2(\xi)={1\over 2} d\xi^i\,d\xi^j\,\partial_i\,Z^{\mathcal N}
\partial_j\,Z^{\mathcal M}\,B_{\mathcal{MN}}(Z)$
is the worldsheet pullback of the NS-NS two--form superfield.

Provided that the superbackground satisfies the IIA supergravity
constraints, the action (\ref{gs}) is invariant under
kappa--symmetry transformations of the superstring coordinates
$Z^{\mathcal M}(\xi)$ which for all known superbranes have the
following generic form
\begin{equation}\label{kalpha}
\delta_\kappa Z^{\mathcal M}\,{\mathcal E}_{\mathcal M}{}^{\underline \alpha}=
{1\over 2}(1+\bar\Gamma)^{\underline \alpha}_{~\underline\beta}\,
\kappa^{\underline\beta}(\xi),\qquad {\underline \alpha}=1,\cdots, 32
\ee
\be\label{kA}
\hskip-2.5cm\delta_\kappa Z^{\mathcal M}\,{\mathcal E}_{\mathcal M}{}^A=0,
\qquad   A=0,1,\cdots,9
\end{equation}
where $\kappa^{\underline\alpha}(\xi)$ is a 32--component spinor
parameter and ${1\over 2}(1+\bar\Gamma)^{\underline
\alpha}_{~\underline\beta}$ is a spinor projection matrix (such that $\bar\Gamma^2=1$) specific
to each type of superbrane.

In the case of the type IIA superstring the matrix $\bar\Gamma$ is
\be\label{gbs}
\bar\Gamma={1\over {2\,\sqrt{-\det{g_{ij}}}}}\,\epsilon^{ij}\,{\mathcal
E}_{i}{}^A\,{\mathcal E}_{j}{}^B\,\Gamma_{AB}\,\Gamma_{11},
\ee
where
\be\label{im}
g_{ij}(\xi)={\mathcal E}_{i}{}^{A}\, {\mathcal
E}_{j}{}^{B}\,\eta_{AB}
\ee
is the induced metric on the worldsheet of the string.

To describe the type IIA superstring in the complete $AdS_4\times
CP^3$ superspace we should just substitute into the above equations
the explicit form of the vector supervielbeins ${\mathcal E}^A$ and
the NS--NS two--form $B_2$ given in eqs. (\ref{sva}) and (\ref{B2}).

Note that kappa--symmetry allows one to eliminate 16 out of 32
fermionic degrees of freedom of the superstring. It can be used to
simplify and reduce the form of the supervielbein pullbacks and, as
a consequence, the form of the string action. For example, one might
be willing, by using kappa--symmetry, to get rid of the 8 fermionic
coordinates $\upsilon^{\alpha i}$ corresponding to the 8 broken
supersymmetries of  $AdS_4\times CP^3$. As a result of such a
partial gauge fixing, one arrives at the superstring action of
\cite{Arutyunov:2008if,Stefanski:2008ik,D'Auria:2008cw}, which can be described
by a sigma--model on the  $OSp(6|4)/U(3)\times SO(1,3)$ supercoset.
However, the kappa--symmetry gauge fixing which completely
eliminates $\upsilon^{\alpha i}$ is only possible when the
kappa--symmetry projector (\ref{kalpha}), (\ref{gbs}) does
\emph{not} commute with the projector ${\mathcal P}_{2}$, eq.
(\ref{p21}), which singles out 8 out of 32 fermionic coordinates.
This is not the case, for example, when the string moves entirely in
the $AdS_4$ space. In this case $[\bar\Gamma,\,{\mathcal P}_2]=0$,
and kappa--symmetry can only eliminate half of the eight
$\upsilon^{\alpha i}$'s. Hence, such configurations of the string in
$AdS_4\times CP^3$ cannot be described by the sigma--model action
based on the $OSp(6|4)/U(3)\times SO(1,3)$ supercoset and one should
use the action (\ref{gs}) in the full superspace.

\subsubsection{Classical Integrability of Green-Schwarz Action in   $AdS_4\times CP^3$  superspace}

The explicit form of the Green-Schwarz action in $AdS_4\times CP^3$
allows for the study of the most general solution   to the  string
equations of motion in this background. Furthermore,  having the
complete action in superspace provides us with a systematic
framework in which to compute quantum corrections to any  classical
string solution. Classical string solutions together with their
quantum corrections play an important role in the AdS/CFT
correspondence, as they correspond to  certain ``long" operators in
the gauge theory (see \emph{e.g.}
\cite{Gubser:2002tv,Frolov:2002av}).

In \cite{Bena:2003wd} it has been shown that the Green-Schwarz
action in $AdS_5\times S^5$ \cite{Metsaev:1998it} is classically
integrable.\footnote{See \cite{Mandal:2002fs} for earlier work
considering the classical integrability of the bosonic
sigma--model.} This can be proven by explicitly constructing a Lax
connection representation of the superstring equations of motion,
such that flatness of the Lax connection ${\cal L}_i$
\be\label{flat}
\partial_{i}{\cal L}_j-\partial_{j}{\cal L}_i-[{\cal L}_i,{\cal L}_j]=0
\ee
implies the  superstring equations of motion. The crucial
ingredients in the construction of the Lax connection are the
Cartan forms in the $AdS_5\times S^5$ coset superspace
$SU(2,2|4)/SO(1,4)\times SO(5)$ and the existence of a $Z_4$
automorphism of  the  $SU(2,2|4)$ algebra. One can then construct
the conserved charges of the integrable model from the holonomy of
the Lax connection (see  \cite{Bena:2003wd}  for more details).

The construction in  \cite{Bena:2003wd} guarantees that any
sigma--model based on a supercoset $G/H$ is classically integrable
as long as the superalgebra $G$ admits a $Z_4$ grading. This general
construction provides a simple diagnostic for determining whether a
large class of supercoset models are classically integrable. In this
paper, however, we have shown that  the complete  $AdS_4\times CP^3$
Type IIA superspace is not given by a coset superspace. This implies
that the technique introduced in \cite{Bena:2003wd} does not
directly apply, as we cannot longer construct a candidate Lax
connection ${\cal L}_i$ from the Cartan forms of the supercoset.
Nevertheless, we can study   the various allowed  motions of the
superstring along submanifolds of the complete  $AdS_4\times CP^3$
Type IIA superspace and analyze whether the equations of motion
governing the allowed motions are classically integrable.

Wherever it is allowed, by partially fixing kappa--symmetry, we can
set to zero the 8 fermionic coordinates $v^{\alpha i}$ which
correspond to the 8 supersymmetries broken by the $AdS_4\times CP^3$
background. This choice selects the submanifold ${\cal M}_{10,24}$
in the complete $AdS_4\times CP^3$ superspace. In this submanifold,
the superstring can  move in the full  $AdS_4\times CP^3$ bosonic
subspace (the string must propagate, however, both in $AdS_4$ and in
$CP^3$ in order to be compatible with the gauge fixing
\cite{Arutyunov:2008if}) but the motion of the string is restricted
to a 24 dimensional fermionic submanifold of the superspace spanned
by the coordinates $\vartheta^{\alpha a'}$. For these classical
configurations, the complete $AdS_4\times CP^3$ superspace found in
this paper reduces to the $OSp(6|4)/U(3)\times SO(1,3)$ coset
superspace already considered in
\cite{Arutyunov:2008if,Stefanski:2008ik,Fre:2008qc,D'Auria:2008cw}.
For this family of classical solutions the Green-Schwarz action can
be completely written down in terms of the Cartan forms of the
$OSp(6|4)/U(3)\times SO(1,3)$  supercoset, very much like for the type IIB
superstring
action on the $AdS_5\times S^5$ coset superspace. Moreover, since
the $OSp(6|4)$ algebra admits a  $Z_4$ automorphism, the
construction in \cite{Bena:2003wd} can be carried over to this case
to show that the classical equations of motion of the superstring in
the subspace ${\cal M}_{10,24}$ of the complete $AdS_4\times CP^3$
superspace is integrable
\cite{Arutyunov:2008if,Stefanski:2008ik}.\footnote{In
\cite{Gromov:2008bz} the algebraic curve characterizing the
classical solutions on this supercoset has been proposed.}

The gauge fixed action with $v^{\alpha i}=0$ is, however,
incompatible with motions of the string \emph{e.g.} purely in the
$AdS_4$ geometry, which constitute an important family of classical
solutions (see \emph{e.g.}  \cite{Gubser:2002tv}). One can study
these motions of the string by considering a submanifold ${\cal
M}_{4,8}$ of the complete  $AdS_4\times CP^3$ superspace. This
submanifold is spanned by the $AdS_4$ bosonic geometry and by the 8
dimensional fermionic space parametrized by the coordinates
$v^{\alpha i}$.\footnote{In this subspace the string worldsheet
scalars $y^{m'}$ are constant and
 $\vartheta^{\alpha a'}$ are covariantly constant (Killing) spinors,
 $D\vartheta=0$ (on the worldsheet).} It follows from our expressions for the
$AdS_4\times CP^3$ superspace, that the submanifold  ${\cal
M}_{4,8}$ can be associated with a ``twisted coset" superspace
$OSp(2|4)/SO(2)\times SO(1,3)$, where the Cartan forms are rotated
by a local Lorentz transformation in $D=11$ superspace, which  was
required to perform the Kaluza-Klein reduction to the $AdS_4\times
CP^3$ superspace. The twisting reflects the fact that the fermionic
coordinates of this superspace correspond to 8 broken
supersymmetries. Thus this superspace does not have superisometries.
The supervielbeins and the Abelian one--form superfield in this
``twisted coset" superspace have the following form (see eqs.
(\ref{sva}) and (\ref{RR1form}))
\bee\label{sva1}
{\mathcal E}^{a}(x,\upsilon) &=&
e^{{1\over3}\phi(\upsilon)}\,\left(e^b(x)-4\upsilon\gamma^b\,{{\sinh^2{{\mathcal
M}/ 2}}\over{\mathcal M}^2}\,D\upsilon\right)\Lambda_b{}^a(\upsilon)
\nonumber\\
&&
+4i\,e^{-{1\over3}\phi(\upsilon)}\,\upsilon\,\ve\gamma^5\,{{\sinh^2{{\mathcal
M}/2}}\over{\mathcal M}^2}\,D\upsilon\, E_7{}^a(\upsilon)\,,
\nonumber\\
\nonumber\\
{\mathcal E}^{\alpha i}(x,\upsilon) &=&
e^{{1\over6}\phi(\upsilon)}\,\left({{\sinh{\mathcal
M}}\over{\mathcal M}}\,D\upsilon\right)^{\beta j}\,S_{\beta
j}{}^{\alpha i}\,(\upsilon) -e^{\phi(\upsilon)}{\mathcal
A}_1(x,\upsilon)\,(i\gamma^{5}\varepsilon\lambda(\upsilon))^{\alpha
i}\,,
\\
\nonumber\\
{\mathcal A}_1(x,\upsilon) &= & e^{-{4\over3}\phi(\upsilon)}\,\left[
\left(e^a(x)-4\upsilon\gamma^a\,{{\sinh^2{{\mathcal
M}/2}}\over{\mathcal M}^2}\,D\upsilon\right)E_{7a}
-4i\upsilon\,\ve\gamma^5\,{{\sinh^2{{\mathcal M}/2}}\over{\mathcal
M}^2}\,D\upsilon\,\Phi(\upsilon)\right]\,,\nonumber
\eee
where
\be\label{dupsilon1}
D\upsilon=\left(d+ie^a(x)\gamma^5\gamma_a-\frac{1}{4}\omega^{ab}(x)\gamma_{ab}\right)\upsilon\,,
\ee
and $e^a(x)$ and $\omega^{ab}(x)$ are the $AdS_4$ vielbeins and
connection respectively. The RR and NS--NS superfields in this
four--dimensional supermanifold have the same form as in
(\ref{f4h3}) but with the D=10 supervielbeins replaced with eqs.
(\ref{sva1}).

For comparison,  let us present the supervielbeins for  the conventional
supercoset \linebreak $OSp(2|4)/SO(2)\times SO(1,3)$
\be\label{sva2}
\begin{aligned}
{\mathcal E}^{a}(x,\upsilon) &=
e^a(x)-4\upsilon\gamma^a\,{{\sinh^2{{\mathcal M}/ 2}}\over{\mathcal
M}^2}\,D\upsilon\,,
\\
\\
{\mathcal E}^{\alpha i}(x,\upsilon) &= \left({{\sinh{\mathcal
M}}\over{\mathcal M}}\,D\upsilon\right)^{\alpha i}\,,
\\
\\
{\mathcal A}_1(x,\upsilon) &=
-4i\upsilon\,\ve\gamma^5\,{{\sinh^2{{\mathcal M}/2}}\over{\mathcal
M}^2}\,D\upsilon\,.
\end{aligned}
\ee

Since the ``twisted $OSp(2|4)/SO(2)\times SO(1,3)$"
supermanifold is not a coset superspace, the criteria used in
\cite{Bena:2003wd} to prove integrability of the classical equations
of motion do not directly apply to this superspace. Therefore, it
remains an important open problem to determine  whether our explicit
form of the Green-Schwarz action when restricted to ${\cal M}_{4,8}$
is classically integrable. The explicit expressions for the
$AdS_4\times CP^3$  supergeometry found in this paper provides a
framework in which this question can be investigated.

Understanding the classical and quantum integrability of the
superstring equations of motion in the $AdS_4\times CP^3$ superspace
also provides an important  clue in determining whether the planar
dilatation operator of the holographic dual ABJM ${\cal N}=6$
Chern--Simons theory is integrable to all orders in the 't Hooft
coupling. Integrability of the two-loop ABJM dilatation operator has
been exhibited in \cite{Minahan:2008hf, Gaiotto:2008cg, Bak:2008cp}
and a conjecture for the all loop Bethe ansatz has been made in
\cite{Gromov:2008qe}. However, unlike for the maximally
supersymmetric $AdS_5/CFT_4$ duality,  the magnon dispersion
relation acquires non-trivial quantum corrections both at strong
coupling as well as in the weak coupling regime
\cite{Nishioka:2008ib, Gaiotto:2008cg, Grignani:2008is, Gromov:2008qe, Ahn:2008aa,
Berenstein:2008dc}, significantly complicating the $AdS_4/CFT_3$
analysis with respect to the case of the  $AdS_5/CFT_4$ duality.
More work is needed to  convincingly argue that the ABJM planar
dilatation operator is exactly integrable. Determining whether  the
ABJM theory is exactly integrable in the planar limit and whether
the Green--Schwarz superstring in the $AdS_4\times CP^3$ superspace
is integrable remain two important problems to resolve in this new
holographic correspondence.

\subsection{Type IIA D--branes}
The action for a Dp--brane  $(p=0,2,4,6,8)$ in a general type IIA
supergravity background
\cite{Cederwall:1996pv,Aganagic:1996pe,Bergshoeff:1996tu} in the
string frame has the form
\begin{equation}\label{DBIstring}
S= -T_p\int\,d^{p+1}\xi\,e^{{ - } {\phi}}\sqrt{-\det(g_{ij}+{\cal
F}_{ij})}+T_p\int\, e^{{\cal F}_2}
\wedge {\mathbb{A}} \vert_{p+1} \,,
\end{equation}
where $T_p$ is the tension of the Dp--brane,
\be\label{imb}
g_{ij}(\xi)={\mathcal E}_{i}{}^{A}\, {\mathcal
E}_{j}{}^{B}\,\eta_{AB}\qquad i,j=0,\ldots,p
\ee
is the induced metric on the Dp--brane worldvolume and
\begin{equation}\label{deltaQstring}
{\cal F}_2 = d{\mathcal V} - {B}_2
\end{equation}
is the field strength of the worldvolume Born--Infeld gauge field
${\mathcal V}_i(\xi)$ extended by the pullback of the NS--NS two--form.
In the second term of eq. (\ref{DBIstring}), the  Wess--Zumino term,
\hbox{ $\vert_{p+1}$ means} that we must pick only the terms which
are   $(p+1)$--forms in the  D-brane worldvolume from the {\it formal}
sum  of the forms of different degrees
$$
e^{{\cal F}_2}=1+ {\cal F}_2+ {1\over 2}\,{\cal F}_2\,{\cal F}_2 +
{1\over {3!}}\,{\cal F}_2\,{\cal F}_2 \,{\cal F}_2+{1\over{4!}
}\,{\cal F}_2\,{\cal F}_2 \,{\cal F}_2\,{\cal F}_2+{1\over
{5!}}\,{\cal F}_2\,{\cal F}_2 \,{\cal F}_2 \,{\cal F}_2\,{\cal
F}_2=\sum_{k=0}^{5}\,{1\over {n!}}\,({\mathcal F}_2)^n\,,
$$
\begin{equation}\label{A}
\mathbb{A}= {\mathcal A}_1 + {\mathcal A}_3 + {\mathcal A}_5 + {\mathcal A}_7 +
{\mathcal A}_9\,=\sum_{n=0}^{4}\,{\mathcal A}_{2n+1}\,,
\end{equation}
where ${\mathcal A}_n$ are the type IIA supergravity RR
superforms and their Hodge duals.

The action (\ref{DBIstring}) is invariant under the kappa--symmetry
transformations (\ref{kalpha})--(\ref{kA}) provided that the
superbackground satisfies the type IIA supergravity constraints and
the Born--Infeld field transforms as follows
\be\label{kV}
\delta_\kappa
\,{\mathcal V}=i_{\delta_\kappa}\,B_2\,\qquad \Rightarrow \qquad \delta_\kappa\,{\mathcal
F}_2=-i_{\delta_\kappa} H_3\,.
\ee
The explicit form of the kappa--symmetry projection matrix $\Gamma$
is given in
\cite{Cederwall:1996pv,Aganagic:1996pe,Bergshoeff:1996tu}.

To describe the Dp--branes in the $AdS_4\times CP^3$ superbackground
one should substitute into the above expressions the explicit form
of the supervielbeins, NS--NS and RR forms given in (\ref{sva}),
(\ref{B2}), (\ref{RR1form}) and (\ref{A3}). As in the superstring
case, one can verify that for  the D0--brane and a D2--brane moving
entirely in the $AdS_4$ space, the corresponding kappa--symmetry
projector  commutes with the projector ${\mathcal P}_2$ (\ref{p21})
which singles out the 8 fermionic coordinates $\upsilon^{\alpha i}$
in superspace. For these configurations,    kappa--symmetry cannot
eliminate all eight $\upsilon^{\alpha i}$'s, but only half of them,
just like for the case of the superstring moving entirely in
$AdS_4$. Therefore, such configurations of D0 and D2-branes cannot
be described by sigma--models based on the supercoset
$OSp(6|4)/U(3)\times SO(1,3)$, and one should use the complete IIA
superspace constructed in this paper.${}^{\ref{fg}}$ In particular,
one should use this complete superspace for studying the
$AdS_4/CFT_3$ correspondence for  the D2--branes placed at the
boundary of $AdS_4$ as well as for the D2--branes corresponding to
vortex loop operators in the boundary field theory
\cite{Drukker:2008jm}.

Other examples of brane configurations for which kappa--symmetry
cannot completely remove the 8 `broken' fermionic coordinates are
D2-- and a D4--branes wrapping the 2--cycle of $CP^3$ associated
with the $CP^3$ K\"ahler form $J$ and moving in $AdS_4$.

In the next sections we shall explain in detail the construction of
the complete type IIA $AdS_4\times CP^3$ superspace which we summarized in
Section \ref{superadscp3}.

\section{Coset space realization of $S^7$ as a fibration over
$CP^3$}\label{s7fibration}

We construct the complete $D=10$ $AdS_4\times CP^3$ superspace by
dimensional reduction of the $D=11$ supercoset $OSp(8|4)/SO(7)\times
SO(1,3)$ whose supervielbeins and superconnection have a fiber
bundle form, generalizing to   superspace the Hopf fibration   form
of the metric and connection of the 7--sphere. So let us start by
reviewing the Hopf fiber bundle structure of the 7--sphere by
considering it as a coset space.

$S^7$ can be realized as the symmetric space $SO(8)/SO(7)$, however
this realization does not provide us directly with the desired Hopf
fibration form of its vielbein and connection. The coset realization
of $S^7$ exhibiting its structure as a Hopf fibration over $CP^3$ is
the coset space $SU(4)\times U(1)\over {SU(3)\times U'(1)}$. Note
that this is not a symmetric space.\footnote{A nice concise review
of geometry of coset spaces the reader may find \emph{e.g.} in
\cite{Mueller-Hoissen}.} On the other hand, $CP^3$ is a symmetric space realized as
the coset $SU(4)\over {SU(3)\times U(1)}$. The isometry group
$SU(4)\times U(1)\simeq SO(6)\times SO(2)$ of the coset $SU(4)\times
U(1)\over {SU(3)\times U'(1)}$ should be considered as a subgroup of
$SO(8)$, $SU(3)$ is a subgroup of $SU(4)$ and $U'(1)$, in the
denominator, is generated as follows. Let $T_2$ be the generator of
$ U(1)\simeq SO(2)$ in the numerator of the coset $SU(4)\times
U(1)\over {SU(3)\times U'(1)}$ and let $T_1$ be the $U(1)$ subgroup
of $SU(4)$ which commutes with $SU(3)$. Then the stability subgroup
$U'(1)$ is generated by
\be\label{T'}
T'={3\over 4}(T_2-T_1)
\ee
and the generator
\be\label{T7} P_7=\frac{1}{4}(3T_1+T_2)\ee
corresponds to the 7th ($U(1)$--fiber) direction of $S^7$. The
inverse expressions are
\be\label{inver}
T_1=P_7-\frac{1}{3}T', \qquad T_2=P_7+T'\,.
\ee
In terms of generators of the $SO(8)$ algebra (See Appendices), the
above generators are
\be\label{T'P7}
T'=-\frac{1}{2}\,J^{a'b'}\,M_{a'b'},\qquad P_7=-M_{78}\,
\ee
where $M_{a'b'}$ are the $SO(6)$ generators and $J_{a'b'}$ are the
components of the K\"ahler form on $CP^3$ satisfying the relations
\begin{equation}\label{jmn1}
J_{{a'}{b'}}=-J_{{b'}{a'}}\,,\qquad
J_{{a'}c'}\,{J^{c'}}_{b'}=-\delta_{{a'}{b'}}\,,\qquad
\epsilon_{{a'}{b'}c'd'e'f'}\,J^{{a'}{b'}}\,J^{c'd'}=8\,J_{e'f'}.
\end{equation}

To get the `fiber bundle form' of the vielbein and connection of the
7--sphere we choose the following coset representative of
$SU(4)\times U(1)\over {SU(3)\times U'(1)}$
\begin{equation}\label{K7}
K=e^{\,y^{m'}P_{m'}}\,e^{\,z T_2}=e^{\,y^{m'}P_{m'}}\,e^{\,z
P_7}\,e^{\,z T'},
\end{equation}
where
$
P_{m'}
$
 are the generators corresponding to the coset
$CP^3={SU(4)\over {SU(3)\times U(1)}}$ parametrized by coordinates
$y^{n'}$ $(n'=1,\cdots,6)$ and $z$ is the $U(1)$ fiber coordinate of
$S^7$ (associated with the generator $P_7$) so that $(y^{n'},\,z)$
are the seven local coordinates on the $S^7$. Note that $e^{\,z T'}$
in (\ref{K7}) plays the role of a compensating local transformation
of the stability subgroup $U'(1)$.

The commutators of $P_{a'}$ close on the $SU(3)$ generators $L_I$
$(I=1,\cdots 8)$ and the $U(1)$ generator $T_1$. Altogether
$P_{a'}$, $L_I$ and $T_1$ form the $SU(4)$ algebra
\be\label{su4}
[P_{a'},P_{b'}]=C_{{a'}{b'}}{}^IL_I+2J_{{a'}{b'}}\,T_1,\quad
[P_{a'},L_I]=C_{{a'}I}{}^{b'}P_{b'},\quad
[P_{a'},T_1]=-\frac{4}{3}\,J_{{a'}{b'}}P^{b'}\,,
\ee
\be\label{su32}
[L_I,L_J]=C_{IJ}{}^KL_K\,,\qquad [L_I,T_1]=0\,,
\ee
where $C_{Ia'b'}$, $C_{IJK}$ and $2\,J_{a'b'}$ are the structure
constants of the $SU(4)$ algebra. In terms of $SO(8)$ generators,
$T_1$ was given in (\ref{inver})-(\ref{T'P7}), and $P_{a'}$ and
$C_{{a'}{b'}}{}^IL_I$ are (See Appendix C for more details)
\be\label{pa'}
P_{a'}=-M_{a'8}+J_{a'}{}^{b'}\,M_{b'7}\,,
\ee
\be\label{LI}
C_{{a'}{b'}}{}^IL_I=(\delta_{a'}^{c'}\,\delta_{b'}^{d'}
+J_{a'}{}^{c'}\,
J_{b'}{}^{d'})M_{c'd'}-\frac{1}{3}\,J_{a'b'}\,J^{c'd'}\,M_{c'd'}.
\ee

The Cartan form $K^{-1}dK$ determines the vielbeins and the
$SU(3)\times U'(1)$ connections on $SU(4)\times U(1)\over
{SU(3)\times U'(1)}$
\bee\label{cfs7}
K^{-1}dK=dy^{{n'}}e_{{n'}}{}^{a'}(y)\,P_{a'}+(dz+dy^{ {n'}}A_{ {n'}}(y))\,P_7\nonumber\\
+dy^{ {n'}}\omega_{ {n'}}{}^I(y)L_I+(dz-\frac{1}{3}\,dy^{ {n'}}A_{
{n'}}(y))\,T'\,,
\eee
where
\be\label{vs7}
{ e}^{{\,\hat a}'}=\left(e^{a'},\,e^7\right),\qquad
e^{a'}=dy^{n'}\,e_{{n'}}{}^{a'}(y)\,,\qquad
e^7=dz+dy^{{n'}}\,A_{{n'}}(y)
\ee
are the $S^7$ vielbeins, with $e^{a'}(y)$ and $A(y)$
being associated with the vielbein and $U(1)$ connection on $CP^3$,
and
\be\label{cs7}
\omega^I=dy^{{n'}}\,\omega_{{n'}}{}^I(y)\,,\qquad
\omega'=dz-\frac{1}{3}\,dy^{{n'}}A_{{n'}}(y)
\ee
are the $SU(3)$ and $U'(1)$ connections  respectively.

With the connections defined as in eq. (\ref{cs7}), the coset space
$SU(4)\times U(1)\over {SU(3)\times U'(1)}$ has  torsion. This is
because its  stability subgroup $U'(1)$ is associated with the
generator $T'$ defined in eq. (\ref{T'}). One can see this analyzing
the Maurer--Cartan equation
\be\label{mc}
d(K^{-1}dK)-(K^{-1}dK)\,(K^{-1}dK)=0\,
\ee
from which follows, in particular, that
\be\label{torsion1}
D\,e^{a'}\equiv
de^{a'}+e^{b'}\,\omega^I\,C_{Ib'}{}^{a'}-e^{b'}\,J_{b'}{}^{a'}\,\omega'=-\,e^{b'}\,e^7\,J_{b'}{}^{a'}\,={1\over
2}\,{e}^{\,{\hat b}'}{e}^{\,{\hat c}'}\,T_{{\hat c}'{\hat
b}'}{}^{a'},
\ee
\be\label{torsion2}
d\,e^7=e^{b'}e^{c'}J_{b'c'}={1\over 2}\,{e}^{\,{\hat
b}'}\,{e}^{\,{\hat c}'}\,T_{{{\hat c}'{\hat b}'} }{}^7\,,
\ee
where $T_{{\hat b}'{\hat c}'}{}^{{\hat a}'}$ (${\hat a}'=(a',7)$
etc.) are the components of the torsion tensor of the coset manifold
$SU(4)\times U(1)\over {SU(3)\times U'(1)}$. To make the geometry on
this manifold torsion--free, as in the standard Riemannian case, we
should redefine its connection as follows
\be\label{Om}
{\Omega}^{\,{\hat a}'{\hat b}'}
=(\Omega^{a'b'},\,\Omega^{a'7})\,,\quad
\ee
{\rm where}
\be\label{Omco}
\Omega^{\,a'b'}=\omega^IC_I{}^{a'b'}-\omega'\,J^{a'b'}=\omega^{a'b'}
-e^7\,J^{a'b'}\,,
\qquad \Omega^{a'7}=-\Omega^{7a'}=e^{b'}J_{b'}{}^{a'}\,
\ee
while
\be\label{cp3om}
\omega^{a'b'}=\omega^IC_I{}^{a'b'}+\frac{4}{3}\,dx^{n'}A_{n'}J^{a'b'}
\ee
is the Riemannian $U(3)$ connection on $CP^3$.

One can show that the curvature of the $SU(4)\times U(1)\over
{SU(3)\times U'(1)}$ coset associated with the connection (\ref{Om}) is
\be\label{R}
d\Omega^{{\hat a}'{\hat b}'}+\Omega^{{\hat a}'}{}_{{\hat
c}'}\,\Omega^{{\hat c}'{\hat b}'}=\,e^{\,{\hat a}'}e^{\,{\hat b}'},
\ee
where the vielbeins $e^{\,{\hat a'}}$ were defined in (\ref{vs7}).
We see that the curvature (\ref{R}) is that of the round $S^7$
sphere.\footnote{We put the radius of $S^7$ and the corresponding
size of $CP^3$ to be one. The $AdS_4$ radius of the $D=11$ and
$D=10$ solution is 1/2 of that of the compact manifold.}  This
completes the demonstration that the Hopf fibration over $CP^3$
associated with the coset space $SU(4)\times U(1)\over {SU(3)\times
U'(1)}$ and endowed with the Riemann connection and curvature is the
7--sphere having   $SO(8)$ isometry, which is enhanced with respect
to the initial $SU(4)\times U(1)$ manifest symmetry of the coset.

The $U(1)$ bundle realization (\ref{vs7}) of the vielbeins of $S^7$
is very convenient for performing the Kaluza--Klein dimensional
reduction of the $AdS_4\times S^7$ solution of $D=11$ supergravity
down to the corresponding $AdS_4\times CP^3$ solution of IIA $D=10$
supergravity \cite{Nilsson:1984bj,Sorokin:1985ap}
\be\label{d11d10}
D=11: \quad e^{\,\hat A}=(e^a,e^{\,{\hat a}'}) \qquad \Rightarrow
\qquad D=10:\quad e^{\,A}=(e^a,e^{a'}),
\ee
where $e^a=dx^{m}e_{m}{}^a(x)$ $(a=0,1,2,3)$ and $x^{m}$
$(m=0,1,2,3)$ are $AdS_4$ vielbeins and coordinates  respectively,
$e^{{\hat a}'}$ are the $S^7$ vielbeins (\ref{vs7}) and
$e^{a'}=dy^{n'}e_{ n'}{}^{a'}(y)$ are the $CP^3$ vielbeins.

For further comparison with the superspace case, it is important to
note that in   the given realization, the components $e_{\hat{
B}}{}^{\hat A}(x,y)$ of the $D=11$ vielbeins of $AdS_4\times S^7$ do
not depend on the $U(1)$ bundle coordinate $z$ and that their
components $e_7{}^a$ and $e_7{}^{a'}$ vanish
\be\label{70}
e_7{}^a=0, \qquad e_7{}^{a'}\,=0\,.
\ee
Such a choice of the vielbein directly corresponds to the
Kaluza--Klein ansatz for the compactification on a circle $S^1$ and
$A_{ m'}(y)$ is associated with the potential of an Abelian
gauge field in the reduced theory.

In our case the field strength of $A_{ m'}(y)$ is the flux
proportional to the K\"ahler form $J_{a'b'}$ on $CP^{3}$
\be\label{fmn}
dA=F_2={1\over 2}\,e^{a'}\,e^{b'}
F_{b'a'}=e^{a'}\,e^{b'}\,J_{a'b'}\,.
\ee
Together with the $F_4$ flux whose non--zero components are along
$AdS_4$, with $F_{abcd}=-6\,\ve_{abcd}$, the $F_2$ flux completes
(the bosonic part of) the compactification of IIA type supergravity on
$AdS_4\times CP^3$.

It should be noted that the Kaluza--Klein condition analogous to
(\ref{70}) is always required in order for the  action and equations
of motion  of the dimensionally reduced theory to have a
conventional gauge structure, describing the interactions of an
Abelian gauge field with gravity. In general, it can always be
achieved by performing an appropriate local Lorentz transformation
of the vielbeins in the original $D+1$--dimensional theory such that
their components with one world index along the compactified
direction and $D$ indices along the reduced D--dimensional tangent
space vanish (as in eq. (\ref{70})).

In the case of the Kaluza--Klein dimensional reduction of the bosonic space
$AdS_4\times S^7$  to ten dimensions, we have arrived at the Kaluza--Klein ansatz
corresponding to the representation of the $S^7$ as
a Hopf fibration over $CP^3$. As we shall see, this is not the
case when the Hopf fibration is lifted to the D=11 supermanifold
$OSp(8|4)/SO(7)\times SO(1,3)$ having 32 fermionic directions. An
additional local Lorentz transformation, which is \emph{not} part of
the $OSp(8|4)$ isometries, will be required to bring the
supervielbeins of this supermanifold to the Kaluza--Klein form, thus
allowing us to perform its dimensional reduction to the $AdS_4 \times
CP^3$ type IIA supergravity solution in $D=10$ superspace with 32
fermionic coordinates.

\section{Lifting the $S^7$ Hopf fibration to $D=11$
superspace}\label{11sgfibration} The superfield descriptions of type
IIA $D=10$ and of $D=11$ supergravity are based on a superspace with
32 fermionic coordinates which in the $AdS_4\times CP^3$ and
$AdS_4\times S^7$ backgrounds can be described by spinors
$\theta^{\alpha\alpha'}$ carrying $AdS_4$ Majorana spinor indices
$(\alpha=1,2,3,4)$ and the indices $(\alpha'=1,\cdots,8)$ of an
8--dimensional spinor representation of $SO(6)$ or $SO(8)$,
respectively. In the $AdS_4\times S^7$ solution of $D=11$
supergravity, $\theta^{\alpha\alpha'}$ are the coordinates of the
coset supermanifold $OSp(8|4)/SO(7)\times SO(1,3)$ associated with
the 32 Grassmann--odd generators $Q_{\alpha\alpha'}$ of  $OSp(8|4)$
(for a detailed description see \emph{e.g.}
\cite{deWit:1998tk,Claus:1998fh}).

On the other hand, the coset supermanifold $OSp(6|4)/U(3)\times
SO(1,3)$ (for its detailed description see \emph{e.g.}
\cite{Arutyunov:2008if,Stefanski:2008ik,Fre:2008qc,D'Auria:2008cw})
is parametrized by ten bosonic coordinates $X^{M}=(x^{m},\,y^{m'})$
corresponding to its bosonic body $AdS_4\times CP^3$ and by 24
fermionic coordinates $\vartheta^{\alpha a'}$, where again
$\alpha=1,2,3,4$ are the $AdS_4$ Majorana spinor indices and
$a'=1,\cdots,6$ are the indices of a 6--dimensional representation
of $SO(6)\simeq SU(4)$. The 24 fermionic coordinates are associated
with the 24 Grassmann--odd generators $Q_{\alpha a'}$ of the
$OSp(6|4)$ algebra.

The 24 generators $Q_{\alpha a'}$ and the corresponding
fermionic coordinates $\vartheta^{\alpha a'}$ can be obtained from
the 32 Grassmann--odd generators $Q_{\alpha\alpha'}$ of $OSp(8|4)$
and the coordinates $\theta^{\alpha\alpha'}$ by acting on the
$SO(8)$ spinor indices with the projection matrix ${\mathcal P}_6$
introduced in
\cite{Nilsson:1984bj} (see \cite{D'Auria:2008cw} and Appendices B and C.2 for more details)
\be\label{p61}
{\mathcal P}_{6}={1\over 8}(6-J)\,,
\ee
where $J$ is the $8\times 8$ symmetric matrix
\be\label{J1}
J=-iJ_{a'b'}\,\gamma^{a'b'}\,\gamma^7 \qquad {\rm such~ that} \qquad
J^2=4J+12\,,
\ee
with $\gamma^{a'}_{\alpha'\beta'}$ $(a'=1,\cdots, 6)$ and
$\gamma^7_{\alpha'\beta'}$ being seven $8\times 8$ gamma matrices
(see Appendix A).

The matrix $J$ has six eigenvalues $-2$ and two eigenvalues $6$,
\emph{i.e.} its diagonalization is given by
\be\label{Jdia1}
J=\hbox{diag}(-2,-2,-2,-2,-2,-2,6,6)\,.
\ee
Therefore, the projector (\ref{p61}) when acting on an
8--dimensional spinor annihilates 2 components and preserves 6 of its components,
while the complementary projector
\be\label{p21}
{\mathcal P}_{2}={1\over 8}(2+J)\,,\qquad {\mathcal P}_2+{\mathcal
P}_6=\mathbf 1
\ee
annihilates 6 and preserves 2 spinor components.

Thus the generators
\be\label{24Q}
({\mathcal P}_6\,Q)_{\alpha\alpha'} \qquad \Longleftrightarrow
\qquad { Q}_{\alpha a'}\,, \qquad a'=1,\cdots, 6
\ee
have 24 non--zero components and can be associated with the 24
Grassmann--odd generators ${Q}_{\alpha a'}$ of $OSp(6|4)$.
Accordingly, the 24 fermionic variables
\be\label{24theta}
({\mathcal P}_6\,\theta)^{\alpha\alpha'} \qquad \Longleftrightarrow
\qquad
\vartheta^{\alpha a'}\,, \qquad a'=1,\cdots, 6
\ee
can be associated with the 24 fermionic coordinates
$\vartheta^{\alpha a'}$ of $OSp(6|4)/U(3)\times SO(1,3)$.

On the other hand, acting on $Q_{\alpha\alpha'}$ with the projector
$\mathcal{P}_2$ (\ref{p21}) one gets 8 generators
\be\label{8Q}
({\mathcal P}_2\,Q)_{\alpha\alpha'} \qquad \Longleftrightarrow
\qquad {\mathcal Q}_{\alpha i},
\qquad i=7,8
\ee
which correspond to the eight supersymmetries broken by the
$AdS_4\times CP^3$ background. The associated 8 fermionic
coordinates of the  type IIA superspace are
\be\label{8theta}
({\mathcal P}_2\,\theta)^{\alpha\alpha'} \qquad \Longleftrightarrow
\qquad
\upsilon^{\alpha i}\,,\qquad i=7,8\,.
\ee
Note that the eight operators $\mathcal{Q}_{\alpha i}$ generate an $OSp(2|4)$
subalgebra of $OSp(8|4)$ (see Appendices B and C.2 for more details)
\bee\label{osp24}
&\{{\mathcal {\mathcal Q}}_{\alpha i},{\mathcal Q}_{\beta
j}\}=-2i\epsilon_{ij}\,\gamma^5_{\alpha\beta}\,T_2
-2\,\delta_{ij}\,(\gamma^a_{\alpha\beta}\,P_a
-i(\gamma^5\gamma^{ab})_{\alpha\beta}\, M_{ab}),\nonumber\\
&\\
&[M_{ab},{\mathcal Q}_{\alpha i}]=-{1\over 2}\,{\mathcal Q}_{\beta
i}\,(\gamma_{ab})^{\beta}{}_\alpha\,,\qquad [P_{a},{\mathcal
Q}_{\alpha i}]=i\,{\mathcal Q}_{\beta
i}\,(\gamma^5\gamma_{a})^{\beta}{}_\alpha\,,
\qquad [T_2,{\mathcal Q}_{\alpha i}]=2\epsilon_{ij}\,{\mathcal Q}_{\alpha j},
\nonumber
\eee
where $T_2$ is the $U(1)\simeq SO(2) $ generator of $SO(8)$ in
$OSp(8|4)$ which commutes with $OSp(6|4)$, so that $OSp(6|4)\times
SO(2)$ is a subgroup of $OSp(8|4)$. Recall that we have introduced
the generator $T_2$ in Section \ref{s7fibration}.

The generators $P_a$ and $M_{ab}$ form the $Sp(4)\simeq Spin(2,3)$
algebra
\be\label{sp41}
[P_a,P_b]=-4M_{ab},\qquad [M_{ab},M_{cd}]=\eta_{ac}\,M_{bd}+
\eta_{bd}\,M_{ac}-\eta_{bc}\,M_{ad}-\eta_{ad}\,M_{bc}\,.
\ee
\be
 [M_{ab},P_c]=\eta_{ac}\,P_b-\eta_{bc}\,P_a\,.
\ee

\subsection{Hopf fibration of the $OSp(6|4)/U(3)\times
SO(1,3)$ supercoset}\label{osp64fibration}

Let us now lift the $OSp(6|4)/U(3)\times SO(1,3)$ solution of IIA
supergravity to $D=11$ by constructing a $U(1)$ bundle over this
supermanifold along the lines of the Hopf fibration of $S^7$
discussed in Section \ref{s7fibration}. This is realized by
constructing a coset superspace
\be\label{hopfosp6}
{{OSp(6|4)\times U(1)}\over{SU(3)\times U'(1)\times SO(1,3)}}\,,
\ee
having 11 bosonic and 24 fermionic directions. In (\ref{hopfosp6})
$U(1)$ is generated by $T_2$ and $U'(1)$ is generated by
$T'=\frac{3}{4}(T_2-T_1)$ (see eq. (\ref{T'})). We take a coset
representative of this superspace in the following form
\be\label{K1124}
K_{_{11,24}}(x,y,z,\vartheta)=K_{_{10,24}}(x,y,\vartheta)\,
e^{\,z\,T_2}\,
\ee
where $K_{_{10,24}}(x,y,\vartheta)$ is a coset representative of
$OSp(6|4)/U(3)\times SO(1,3)$ which can be taken in any convenient
form, \emph{e.g.} in one of those considered in
\cite{Arutyunov:2008if,Stefanski:2008ik,Fre:2008qc,D'Auria:2008cw} (or Appendix $A$).

The supervielbeins and superconnections of the supercoset
(\ref{hopfosp6}) are encoded in the \linebreak ${OSp(6|4)\times
U(1)}$ Cartan form
\be\label{c1124}
K^{-1}_{_{11,24}}\,d\,K_{_{11,24}}=K^{-1}_{_{10,24}}\,d\,K_{_{10,24}}+dz\,T_2\,,
\ee
where the $OSp(6|4)$ Cartan form
\bee\label{c1024}
K^{-1}_{_{10,24}}\,d\,K_{_{10,24}}=E^{a}(x,y,\vartheta)\,P_a+E^{a'}(x,y,\vartheta)\,P_{a'}+E^{\alpha
a'}(x,y,\vartheta)\,Q_{\alpha a'}\nonumber\\
\\
+{1\over
2}\Omega^{ab}(x,y,\vartheta)\,M_{ab}+\Omega^{I}(x,y,\vartheta)\,L_{I}+A(x,y,\vartheta)\,T_1\nonumber
\eee
contains the supervielbeins and superconnections of
$OSp(6|4)/U(3)\times SO(1,3)$ whose explicit form can be found in
\cite{Arutyunov:2008if,Stefanski:2008ik,Fre:2008qc,D'Auria:2008cw} (or Appendix $A$).
The $SU(3)\times U(1)$ generators $L_I$ and $T_1$ have been
introduced in Section \ref{s7fibration}.

Now, as in the case of the 7--sphere, eq. (\ref{cfs7}), we single
out proper supervielbeins and superconnections of the supercoset
(\ref{hopfosp6}) as follows
\bee\label{c11241}
&K^{-1}_{_{11,24}}\,d\,K_{_{11,24}}=E^{a}(x,y,\vartheta)\,P_a+E^{a'}(x,y,\vartheta)\,P_{a'}
+\left(dz+A(x,y,\vartheta)\right)\,P_7 +E^{\alpha
{a'}}(x,y,\vartheta)\,Q_{\alpha {a'}}\nonumber\\
&\\
&+{1\over 2}\Omega^{ab}(x,y,\vartheta)\,M_{ab}
+\Omega^{I}(x,y,\vartheta)\,L_{I}+(dz-\frac{1}{3}\,A(x,y,\vartheta))\,T'\nonumber\,.
\eee
Given that
\be\label{sc}
Z^{\tilde{\mathcal M}}=(x^{m},y^{ m'},\vartheta^{\alpha a'})
\ee
are the supercoordinates parametrizing $OSp(6|4)/U(3)\times SO(1,3)$ and that $z$ is the coordinate of the
Hopf fiber,
the 11 bosonic and 24 fermionic supervielbeins are given by
\bee\label{E11}
&E^{\hat A}=(E^a,\,E^{{\hat a}'})\,, \quad E^a=dZ^{\tilde{\mathcal
M}}\,E_{\tilde{\mathcal M}}{}^a(x,y,\vartheta)\,, \quad E^{\,{\hat
a}'}=dZ^{\tilde{\mathcal M}}\,E_{\tilde{\mathcal M}}{}^{{\hat
a}'}(x,y,\vartheta)=(E^{a'}\,,E^7)\,,
\nonumber \\
\\
&E^7=dz+dZ^{\tilde{\mathcal M}}\,A_{\tilde{\mathcal M}}(x,y,\vartheta)\nonumber
\eee
while  the 24 fermionic supervielbeins are
\be\label{E24}
E^{\alpha a'}=dZ^{\tilde{\mathcal M}}\,E_{\tilde{\mathcal
M}}{}^{\alpha a'}(x,y,\vartheta)\,.
\ee
The connections of the stability group $U(3)\times SO(1,3)$ are
given in the last line of (\ref{c11241}).

We see that the components of the supervielbeins and connections do
not depend on the 11th coordinate $z$, which appears only in  the
differential of the $E^7$ vielbein. Moreover, the supervielbein components
$dz\,E^{\cal A}_7=(dz\,E_7{}^a,\, dz\,E_7{}^{a'},\,dz\,
E_7{}^{\alpha a'})$ are all zero. Thus, the realization of the coset
supermanifold (\ref{hopfosp6}) considered above has a Hopf fibration
structure generalizing that of the 7--sphere. The dimensional
reduction of this supermanifold to $D=10$ is then straightforward. One
must  just project it orthogonally to the $U(1)$ fiber direction, \emph{i.e.}
to pick $E^a$, $E^{a'}$ and $E^{\alpha a'}$ as the supervielbeins of
the $D=10$ superspace and to consider $dZ^{\tilde{\mathcal
M}}\,A_{\tilde{\mathcal M}}(x,y,\vartheta)$ as the RR one--form
potential of the type IIA supergravity theory. Note that in this reduced type IIA superspace solution,
the dilaton superfield is constant and the dilatino superfield  vanishes.

The difference with respect to the purely bosonic case is that whereas the
$S^7$ fibration has an enhanced $SO(8)$ isometry, the isometry
supergroup of the supermanifold (\ref{hopfosp6}) is still
$OSp(6|4)\times U(1)$, since $SO(8)$ is not its subgroup. The
extension to $SO(8)$ and, hence, to $OSp(8|4)$ requires the
introduction of 8 additional Grassmann--odd generators.

On the other hand, it can be directly verified that the $D=11$
superspace with 24 fermionic directions considered above is a
solution of superfield constraints of $D=11$ supergravity (and,
hence, of its equations of motion). It thus provides a description
of the maximally supersymmetric $AdS_4\times S^7$ solution in a
reduced superspace which can be regarded as a sub-superspace of
$OSp(8|4)/SO(7)\times SO(1,3)$.

\subsection{$U(1)$ bundle structure of the $OSp(8|4)/SO(7)\times SO(1,3)$ supercoset}
Let us now extend the supercoset (\ref{hopfosp6}) to the full
$OSp(8|4)/SO(7)\times SO(1,3)$ supercoset. This is achieved by taking the
following group element of $OSp(8|4)$ as the coset representative of
$OSp(8|4)/SO(7)\times SO(1,3)$
\be\label{hopfosp8}
K_{_{11,32}}(x,y,z,\theta)=K_{_{11,24}}(x,y,z,\vartheta)\,e^{\,\upsilon^{\alpha
i}\,{\mathcal Q}_{\alpha i}}=K_{_{10,24}}(x,y,\vartheta)\,
e^{\,z\,T_2}\,\,e^{\,\upsilon^{\alpha i}\,{\mathcal Q}_{\alpha i}}\,,
\ee
where $K_{_{11,24}}(x,y,z,\vartheta)$ is the same coset
representative as in (\ref{K1124}) and $\theta=(\vartheta,\upsilon)$
are the 32--component fermionic coordinates which, using the
projectors (\ref{p61}) and (\ref{p21}),  split into 24--component
$\vartheta$'s and 8--component $\upsilon$'s. Note that the group
element $e^{\,\upsilon^{\alpha i}\,{\mathcal Q}_{\alpha i}}$ can be
regarded as the representative of the purely fermionic supercoset
${{OSp(2|4)}\over{SO(2)\times SO(2,3)}}$.

The $OSp(8|4)$--valued Cartan form constructed with
(\ref{hopfosp8}) is
\bee\label{c1132}
&K^{-1}_{_{11,32}}\,dK_{_{11,32}}=e^{\,-\upsilon\,{\mathcal Q}}\,
(K^{-1}_{_{11,24}}\,d\,K_{_{11,24}})\,e^{\,\upsilon\,{\mathcal Q}}+e^{\,-\upsilon\,{\mathcal
Q}}\, d\,e^{\,\upsilon\,{\mathcal Q}}\nonumber\\
&\\
&=e^{\,-\upsilon\,{\mathcal
Q}}\,(K^{-1}_{_{10,24}}\,d\,K_{_{10,24}})\,e^{\,\upsilon\,{\mathcal
Q}}+dz\, e^{\,-\upsilon\,{\mathcal
Q}}\,T_2\,e^{\,\upsilon\,{\mathcal Q}}+e^{\,-\upsilon\,{\mathcal
Q}}\, d\,e^{\,\upsilon\,{\mathcal Q}}\,\nonumber
\eee
or, using the commutation relations (\ref{osp24}) and the form of
$K^{-1}_{_{10,24}}\,d\,K_{_{10,24}}$ (\ref{c1024})
\bee\label{car1132}
&\hspace{-10pt}K^{-1}_{_{11,32}}\,dK_{_{11,32}}
=E^{a}(x,y,\vartheta)\,e^{\,-\upsilon\,{\mathcal
Q}}\,P_a\,e^{\,\upsilon\,{\mathcal Q}}+E^{\alpha
a'}(x,y,\vartheta)\,e^{\,-\upsilon\,{\mathcal Q}}\,Q_{\alpha
a'}\,e^{\,\upsilon\,{\mathcal Q}}
+E^{a'}(x,y,\vartheta)\,P_{a'}\nonumber\\
&\nonumber\\
&+{1\over 2}\Omega^{ab}(x,y,\vartheta)\,e^{\,-\upsilon\,{\mathcal
Q}}\,M_{ab}\,e^{\,\upsilon\,{\mathcal
Q}}+\Omega^{I}(x,y,\vartheta)\,L_{I}+A(x,y,\vartheta)\,T_1\\
&\nonumber\\
&+dz\, e^{\,-\upsilon\,{\mathcal Q}}\,T_2\,e^{\,\upsilon\,{\mathcal
Q}}+e^{\,-\upsilon\,{\mathcal Q}}\, d\,e^{\,\upsilon\,{\mathcal
Q}}\,.\nonumber
\eee
Note that the supervielbein and connection terms in (\ref{car1132})
corresponding to the $SU(4)$ generators $P_{a'}$, $L_I$ and $T_1$ do
not receive contributions from $\upsilon^{\alpha i}$, since these generators commute
with ${\mathcal Q_{\alpha i}}$.

Furthermore, we can expand the Cartan form (\ref{car1132}) in the
basis of the $OSp(8|4)$ generators. The expansion   contains
generators along the  $AdS_4$,
$CP^3$ and  $z$ directions, along the generators of their stability group
$SO(1,3)\times SU(3)\times U'(1)$ and the rest. It is given by
\bee\label{cartan1132}
K^{-1}_{_{11,32}}\,dK_{_{11,32}}
=E_{_{11,32}}^a\,P_a+E_{_{11,32}}^{a'}\,P_{a'}+E_{_{11,32}}^7\,P_7+E_{_{11,32}}^{\alpha
i}\,{\mathcal Q_{\alpha i}}+E_{_{11,32}}^{\alpha a'}\,{ Q_{\alpha
a'}}\nonumber\\
\\
+{1\over 2}\Omega_{_{11,32}}^{ab}\,M_{ab}+
\Omega_{_{10,24}}^{I}\,L_I+\Omega'_{_{11,32}}\,T'+\tilde\Omega_{_{11,32}}^{a'i}\,{\tilde
M}_{a'i},
\nonumber
\eee
where, we remind the reader that  $P_7$ and $T'$ were defined in
(\ref{T'}) and (\ref{T7}), while
\be\label{mai+pa}
{\tilde M}_{a'i}\quad \Leftrightarrow\quad
\frac{1}{4}\,{\mathcal P}_6\gamma^{\tilde a'\tilde b'}{\mathcal P}_2\,M_{\tilde a'\tilde
b'}+\frac{i}{2}\,{\mathcal P}_6\gamma^{a'}{\mathcal P}_2\,\,P_{a'},
\ee
with $M_{\tilde a'\tilde b'}$ being the generators of $SO(8)$ (see
Appendix B). ${\tilde M}_{a'i}$ are the generators which complete
the $SO(6)\times SO(2)$ algebra to $SO(8)$. They differ from the
generators $M_{a'i}$ introduced in Appendix B, eqs.
(\ref{q6q2})--(\ref{qmai}), by the shift along the $CP^3$
translations generated by $P_{a'}=-M_{a'8}+J_{a'}{}^{b'}\,M_{b'7}$.
The reason for this redefinition is that the commutators of the
generators $M_{a'i}$, defined in eqs. (\ref{ai}), produces the
generators of the $SO(6)\times SO(2)$ subgroup of the $SO(8)$ group,
and, in particular the $CP^3$ coset generators $P_{a'}$. Thus,
$M_{a'i}$ themselves cannot be regarded as generators belonging to
the structure group $SO(7)$ of the 7--sphere. The commutators of the
$SO(7)$ generators should not produce the translations along $S^7$.
Therefore, to make $M_{a'i}$ part of $SO(7)$ one must redefine them
as in eq. (\ref{mai+pa}). This redefinition results in the
appearance of the additional (second) term in the expression for the
supervielbein $E^{a'}_{11,32}$ in eq. (\ref{upsilonfunctions})
below.

All functions of $\upsilon^{\alpha i}$ in
(\ref{cartan1132}) can be explicitly computed using the commutation
relations (\ref{osp24}), (\ref{q6q2}) and (\ref{qmai}) and applying
the method described
\emph{e.g.} in \cite{Kallosh:1998zx,Metsaev:1998hf,deWit:1998tk,Claus:1998fh}.
The  supervielbeins we get are
\be\label{upsilonfunctions}
\begin{aligned}
E_{_{11,32}}^a&=E_{_{10,24}}^a-4\upsilon\gamma^a\,{{\sinh^2{{\mathcal
M}/ 2}}\over{\mathcal M}^2}\,D\upsilon\,
+dz\,E_7{}^a(\upsilon)\,,\\
E_{_{11,32}}^{a'}&=E_{_{10,24}}^{a'}-2\upsilon\,{{\sinh m}\over m}\gamma^{a'}\gamma^5\,E_{_{10,24}}\,,\\
E_{_{11,32}}^7&=dz\,\Phi(\upsilon)+A_{_{10,24}}-4i\upsilon\,\ve\gamma^5\,{{\sinh^2{{\mathcal
M}/2}}\over{\mathcal M}^2}\,D\upsilon\,, \\
E_{_{11,32}}^{\alpha i}&=\left({{\sinh{\mathcal M}}\over{\mathcal
M}}\,(D\upsilon-2dz\,\ve\upsilon)\right)^{\alpha i}\,,
\\
E_{_{11,32}}^{\alpha a'}&=E_{_{10,24}}^{\alpha
a'}-8E_{_{10,24}}^{\beta
a'}\left(i\gamma^5\,\upsilon\,{{\sinh^2{{m}/2}}\over{m}^2}\right)_{\beta
i}\upsilon^{\alpha i}\,,
\end{aligned}
\ee
the $SO(1,3)$ connection is
\be\label{upsilonfunctions1}
\Omega_{_{11,32}}^{ab}= \Omega_{_{10,24}}^{ab}
+8i\upsilon\gamma^{ab}\gamma^5\,{{\sinh^2{{\mathcal M}/
2}}\over{\mathcal
M}^2}\,\left(D\upsilon-2dz\,\ve\upsilon\right)\,,\\
\ee
the one--form $\tilde\Omega^{a'i}$ is
\be\label{omegaa'i}
\tilde\Omega_{_{11,32}}^{a'i}=4E_{_{10,24}}^{\alpha a'}\,
\left(i\gamma^5\,\upsilon\,{{\sinh m}\over m}\right)_{\alpha}^{i}
\ee
and the one-form $\Omega'_{_{11,32}}$ is
\be\label{A1132}
\Omega'_{_{11,32}}=dz\,\Phi(\upsilon)-\frac{1}{3}\,A_{_{10,24}}-4i\upsilon\,\ve\gamma^5\,{{\sinh^2{{\mathcal
M}/ 2}}\over{\mathcal M}^2}\,D\upsilon\,.
\ee
The $SO(7)$ connection in the considered realization of the
supercoset $OSp(8|4)/SO(7)\times SO(1,3)$ can be computed from
(\ref{cartan1132}) and has the form
\bee\label{so7connection}
\frac{1}{2}\,\Omega_{_{11,32}}^{a'b'}\,M_{a'b'}
+\Omega_{_{11,32}}^{a'7}\,M_{a'7}&=&(E_{_{11,32}}^{b'}+4\upsilon\,{{\sinh
m}\over
m}\gamma^{b'}\gamma^5\,E_{_{10,24}})\,J_{b'}{}^{a'}\,M_{a'7}\nonumber\\
\\
&&\hspace{-100pt}+{\frac{1}{2}}\,(\Omega_{_{10,24}}^{a'b'}
-E^7_{_{11,32}}\,J^{a'b'} -2i\upsilon\,{{\sinh m}\over
m}\gamma^{a'b'}\gamma^5\,E_{_{10,24}})\,M_{a'b'}.\nonumber
\eee
 The functions appearing in
(\ref{upsilonfunctions})--(\ref{A1132}) are defined as\footnote{Note
that only positive even powers of $\mathcal M$ and $m$ appear in the
above expressions when they are expanded.}
\be\label{U}
\begin{aligned}
({\mathcal M}^2)^{\alpha i}{}_{\beta j}&=4i(\ve\upsilon)^{\alpha
i}(\upsilon\ve\gamma^5)_{\beta
j}-2i(\gamma^5\gamma^a\upsilon)^{\alpha i}(\upsilon\gamma_a)_{\beta
j}-i(\gamma^{ab}\upsilon)^{\alpha
i}(\upsilon\gamma_{ab}\gamma^5)_{\beta j}\,,
\\
\\
(m^2)^{ij}&=-4i\upsilon^{ i}\,\gamma^5\,\upsilon^{j}\,,
\end{aligned}
\ee
\be\label{phiE7}\begin{aligned}
\hskip-6.6cm E_7{}^a(\upsilon)&=8\,\upsilon\gamma^a\,{{\sinh^2{{\mathcal M}/
2}}\over{\mathcal M}^2}\,\varepsilon\,{\upsilon}\,,
\\
\hskip-6.6cm \Phi(\upsilon)&= 1+8i\,\upsilon\,\ve\gamma^5\,{{\sinh^2{{\mathcal
M}/2}}\over{\mathcal M}^2}\,\ve\upsilon
\end{aligned}
\ee
and
\bee\label{D}
D\upsilon=\left(d+iE^a_{_{10,24}}\gamma^5\gamma_a-\frac{1}{4}\Omega^{ab}_{_{10,24}}\gamma_{ab}\right)\upsilon\,.
\eee
All quantities in (\ref{upsilonfunctions})--(\ref{D}) labeled as
$E_{10,24}$, $\Omega_{10,24}$ \emph{etc.} are the ones which
describe the supercoset $OSp(6|4)/U(3)\times SO(1,3)$ and are
explicitly known (see \emph{e.g.}
\cite{Kallosh:1998zx,Arutyunov:2008if,Stefanski:2008ik,Fre:2008qc,D'Auria:2008cw} and Appendix $A$).

Analyzing eqs. (\ref{car1132})--(\ref{D}) we observe, in particular,
that due to the multiplication by $e^{\,\upsilon\,Q}$ in
(\ref{car1132}), the $AdS_4$ supervielbeins and the $SO(1,3)$
superconnections (\ref{c1024}) corresponding to the supercoset
$OSp(6|4)/U(3)\times SO(1,3)$ acquire non-trivial dependence on the
8 additional fermionic variables $\upsilon^{\alpha i}$. In the first
line of (\ref{car1132}) and in (\ref{cartan1132}) there are also
terms with components of the superconnection corresponding to the
generators (\ref{mai+pa}) which extend the $SO(6)\times SO(2)$
algebra to $SO(8)$ because of the non--trivial anti--commutators of
the 24 supersymmetry generators $Q_{\alpha a'}$ with the 8
supersymmetry generators ${\mathcal Q}_{\alpha i}$ (eqs.
(\ref{q6q2})--(\ref{qmai})).

We also observe that, in contrast to the  cases discussed in
Sections
\ref{s7fibration} and \ref{osp64fibration}, the $U(1)$--bundle realization of the
$OSp(8|4)/SO(7)\times SO(1,3)$ supercoset geometry in
(\ref{cartan1132}) does not allow for its direct dimensional
reduction to a $D=10$ superspace because of the presence of the term
$dz\,E_7{}^a(\upsilon)$. This term contributes to the components of
the supervielbein along the directions tangent to $AdS_4$ and has a
`leg' along the compactified direction parametrized by the
$z$--coordinate.\footnote{A somewhat amusing remark is that the term
$dz\,E_7{}^a(\upsilon)$, in a certain sense, `mixes' the $AdS_4$
geometry with the $U(1)$ fiber direction of the $S^7$. On the other
hand, the more `natural' terms like $dz\,E_7{}^{a'}(\upsilon)$ along
the $CP^3$ tangent space, which would mix the Hopf fiber direction
with $CP^3$,  are absent. They would correspond to some vielbein
components on the $S^7$.}  As we discussed in the end of Section
\ref{s7fibration}, to perform the Kaluza--Klein dimensional
reduction such components of the (super)vielbein must be put to
zero.

From the supervielbeins in (\ref{upsilonfunctions}) we can also
construct the supergeometry corresponding to the superspace with
$AdS_4\times S^7/Z_k$ bosonic body, a background of eleven
dimensional supergravity which preserves 24 supersymmetries (for
$k>2$) and is the near horizon geometry of N M2–-branes probing the
$C^4/Z_k$ singularity. Geometrically, this superspace is obtained by
orbifolding the $OSp(8|4)/SO(7)\times SO(1,3)$ supercoset geometry
by $Z_k\subset U(1)$, where $U(1)$ is the commutant of $SU(4)$ in
$SO(8)$. The corresponding supervielbeins are simply obtained from
those in (\ref{upsilonfunctions}) by replacing $z\rightarrow z/k$.

\subsection{Hopf fibration form of the $OSp(8|4)/SO(7)\times SO(1,3)$
geometry and its reduction to type IIA superspace}\label{lorentzrot}

To eliminate the term $dz\,E_7{}^a(\upsilon)$ from the
$OSp(8|4)/SO(7)\times SO(1,3)$ supervielbein we should perform an
appropriate local Lorentz rotation in the 5--plane $(E^a,E^7)$
tangential to $AdS_4\times S^1$, where $S^1$ is the $U(1)$ fiber
direction in  $S^7$. Obviously, such a transformation is not an
isometry of the coset supermanifold $OSp(8|4)/SO(7)\times SO(1,3)$
and should therefore be regarded simply as a change of local
frame. Upon this Lorentz transformation we shall get the $D=11$
supervielbeins in a form which will allow us to directly identify the
corresponding $D=10$ supervielbeins, the RR one--form gauge
superfield and the dilaton superfield of type IIA supergravity.

Let ${E}^{\hat A}=(E^a,\,E^{a'},\,E^7)$ be the 11 bosonic components of the
$OSp(8|4)/SO(7)\times SO(1,3)$ supervielbein given in (\ref{upsilonfunctions}). To eliminate the $dz\,E_7{}^a(\upsilon)$
component of $E^a$ we perform the following Lorentz transformation
\be\label{lorentz}
\underline{\mathcal E}^a=E^b\,\Lambda_b{}^a(\upsilon)+E^7\,\Lambda_7{}^a(\upsilon)\,,\qquad
\underline{\mathcal E}^7=E^b\,\Lambda_b{}^7(\upsilon)+E^7\,\Lambda_7{}^7(\upsilon)\,,
\ee
where the parameters $\Lambda_{\hat b}{}^{\hat a}(\upsilon)$ $(\hat
a=(a,7)=0,1,2,3,7)$ depend on the 8 fermionic coordinates
$\upsilon^{\alpha i}$ and satisfy the 5--dimensional Lorentz group
orthogonality conditions
\be\label{oc}
\Lambda_{\hat a}{}^{\hat c}\,\Lambda_{\hat b}{}^{\hat d}\,\eta_{\hat c\hat
d}=\eta_{\hat a\hat b},
\ee
or in components
\be\label{occ1}
\Lambda_{a}{}^{c}\,\Lambda_{b}{}^{d}\,\eta_{cd}+\Lambda_{a}{}^{7}\,\Lambda_{b}{}^{7}=\eta_{a
b},\quad
\Lambda_{7}{}^{c}\,\Lambda_{7}{}^{d}\,\eta_{cd}+(\Lambda_{7}{}^{7})^2=1,\quad
\Lambda_{7}{}^{c}\,\Lambda_{a}{}^{d}\,\eta_{cd}+\Lambda_{7}{}^{7}\Lambda_{a}{}^{7}=0
\ee
and
\be\label{occ2}
\Lambda_{a}{}^{c}\,\Lambda_{b}{}^{d}\,\eta^{ab}+\Lambda_{7}{}^{c}\,\Lambda_{7}{}^{d}=\eta^{c
d},\quad
\Lambda_{c}{}^{7}\,\Lambda_{d}{}^{7}\,\eta^{cd}+(\Lambda_{7}{}^{7})^2=1,\quad
\Lambda_{c}{}^{7}\,\Lambda_{d}{}^{a}\,\eta^{cd}+\Lambda_{7}{}^{7}\Lambda_{7}{}^{a}=0\,.
\ee

In addition $\Lambda_{\hat b}{}^{\hat a}(\upsilon)$ is determined
by the requirement that the $\underline{\mathcal E}_7{}^a$ component of the transformed
supervielbein vanishes and that at $\upsilon=0$ it reduces to the unit matrix
\be\label{7a}
\Lambda_{\hat b}{}^{\hat a}(\upsilon)|_{\upsilon=0}=\delta_{\hat b}{}^{\hat
a}\,,
\qquad
\underline{\mathcal E}_7{}^a=E_7{}^b\,\Lambda_b{}^a+\Phi\,\Lambda_7{}^a=0\,,
\ee
where $\Phi(\upsilon):=E_7{}^7, \,\,\Phi(0)=1$ (see eq.
(\ref{phiE7})). From eq. (\ref{7a}) we find that
\be\label{L7a}
\Lambda_7{}^a(\upsilon)=-{1\over
\Phi(\upsilon)}\,E_7{}^b(\upsilon)\,\Lambda_b{}^a(\upsilon)\,.
\ee
Then, solving the orthogonality conditions (\ref{occ1}) and
(\ref{occ2}) we find the expressions for the parameters of the
Lorentz transformation in terms of $E_7{}^a(\upsilon)$ and
$\Phi(\upsilon)$
\be\label{L77}
\Lambda_7{}^7={\Phi\over{\sqrt{\Phi^2+E^2}}}\,,
\ee
\be\label{La7}
\Lambda_a{}^7={E_{7\,a}\over{\sqrt{\Phi^2+E^2}}}\,,
\ee
where $E^2\equiv E_7{}^a\,E_7{}^b\,\eta_{ab}$, and
\bee\label{Lab}
\Lambda_{a}{}^{c}\,\Lambda_{b}{}^{d}\,\eta_{cd}&=&\eta_{ab}
-{{E_{7\,a}\,E_{7\,b}}\over{{\Phi^2+E^2}}}\,,
\nonumber\\
\\
\Lambda_a{}^{b}=\delta_a{}^{b}-{E_{7a}}\,E_7{}^b\,
{{\sqrt{\Phi^2+E^2}-\Phi}\over{E^2}\,\sqrt{\Phi^2+E^2}}&\Rightarrow&
\det\Lambda_a{}^b=\frac{\Phi}{\sqrt{\Phi^2+E^2}}\,.\nonumber
\eee
Finally eq. (\ref{L7a}) can be rewritten as
\be\label{L7a1}
\Lambda_7{}^a=-{{E_7{}^a}\over{\sqrt{\Phi^2+E^2}}}\,.
\ee

One can notice that $\Lambda_{\hat b}{}^{\hat a}$ depend only on the
vector parameter ${1\over\Phi}\,{E_7{}^a}$ and thus can be regarded
as a kind of ``Lorentz boost" along the $S^7$ fiber direction.

The following ten components of the Lorentz transformed $D=11$
supervielbeins
\bee\label{IIAE1}
&\underline{\mathcal E}^A(x,y,\vartheta,\upsilon)=(\underline{\mathcal E}^a,\,\underline{\mathcal E}^{a'})\,,\qquad
A=0,1,\cdots, 9;\quad a=0,1,2,3; \quad a'=1,\cdots, 6\nonumber\\
&\\
&\underline{\mathcal E}^a=E^b\,\Lambda_b{}^a(\upsilon)+E^7\,\Lambda_7{}^a(\upsilon)\,,\qquad
\underline{\mathcal E}^{a'}=E^{a'}\nonumber
\eee
form an appropriate bosonic supervielbein of the complete
(32\,-\,$\theta$) superfield solution of type IIA supergravity
corresponding to the $AdS_4\times CP^3$ vacuum. The IIA dilaton
superfield is
\be\label{dilaton}
e^{{2\over
3}\phi(\upsilon)}=\Phi\,\Lambda_7{}^7+E_7{}^a\,\Lambda_a{}^7
=\sqrt{\Phi^2+E_7{}^a\,E_7{}^b\,\eta_{ab}}\,.
\ee
One can notice that the dilaton superfield of this type IIA solution
depends only on the eight fermionic coordinates $\upsilon^{\alpha i}$
which correspond to the broken supersymmetries of the $AdS_4\times
CP^3$ background.

In addition to the Lorentz rotation of the vector supervielbeins, we
should also perform a corresponding Lorentz rotation of the components
of the connections  and of the spinor supervielbeins $E^{\alpha a'}$
and $E^{\alpha i}$. In particular, the Lorentz rotation of the
connection components $\Omega^{\,a'7}$ will produce a ``mixed"
$AdS_4$--$CP^3$ term
$\Omega^{\,a'a}=\Omega^{\,a'7}{}\Lambda_7{}^{a}$ which transforms as
a tensor under $U(3)\times SO(1,3)$ and hence can be absorbed into a
redefined torsion of the type IIA superspace.

As far as the Lorentz rotation of the spinor supervielbeins is
concerned, it is worth noting that the Lorentz rotation of the
spinors associated with (\ref{oc})--(\ref{occ2}) is generated by the
gamma--matrices
$\Gamma^a\Gamma^{11}=\gamma^a\,\gamma^5\otimes\gamma^7$ which
commute with the projectors (\ref{p61}) and (\ref{p21}) and thus
does not mix the 24  and 8--component spinors. The explicit form of the
Lorentz rotation acting on spinors, $S_{\underline
\alpha}{}^{\underline\beta}(\upsilon)$
($\underline\alpha=(\alpha\alpha')$), can be derived using the well
known relations between the vector and spinor representations of the
Lorentz group
\be\label{SS}
S^{-1}\,\Gamma^{\hat a}\,S=\Gamma^{\hat b}\,\Lambda_{\hat b}{}^{\hat
a}\,,
\qquad
S_{\underline \alpha}{}^{\underline\gamma}\,\,S_{\underline
\beta}{}^{\underline\delta}\,\,{\mathcal
C}_{\underline{\gamma}\underline\delta} ={\mathcal
C}_{\underline{\alpha}\underline\beta}\,,
\ee
where $\Gamma^{\hat a}=(\Gamma^{a},\Gamma^{11})$ are $32\times 32$
gamma--matrices defined in (\ref{Gamma10}) and ${\mathcal
C}=C\otimes C'$ is the corresponding charge conjugation matrix.

Since the Lorentz transformation giving rise to supervielbeins and
connections compatible with the KK ansatz corresponds to a Lorentz
rotation with the ``velocity" parameter ${\rm w}^a=E_7{}^a/\Phi$,
the corresponding matrix acting on the fermions (\ref{SS})  is given
by
\be\label{fermionicboost}
\begin{aligned}
S&=\exp(-\frac{1}{2}\,\frac{{\rm w}^a}{|{\rm
w}|}\,{\Gamma_a}\Gamma_{11}\,\tan^{-1}|{\rm w}|) \hspace{100pt}
({\rm w}^a=E_7{}^a/\Phi)
\\
\\
&=2^{-1/2}(1+{\rm w}^2)^{-1/4}\left(
\sqrt{\sqrt{1+{\rm w}^2}+1}
-\frac{{\rm w}^a}{|{\rm w}|}\,\Gamma_a\Gamma_{11}\sqrt{\sqrt{1+{\rm w}^2}-1}
\,\right)
\,.
\end{aligned}
\ee

Performing the Lorentz rotation described above, the $D=11$
supervielbeins (upon a Weyl rescaling) acquire a form which is
suitable for the dimensional reduction to $D=10$ superspace in the
string frame
\cite{Duff:1987bx,Howe:2004ib}:
\bee\label{KKform}
&\underline{\mathcal E}^{\hat A}=(e^{-{1\over 3}\phi}\,{\mathcal E}^A,\,\underline{\mathcal E}^{11})\,,\qquad
\underline{\mathcal E}^{11}=e^{{2\over 3}\phi}\,(dz+{\mathcal A}_1),\nonumber\\
&\\
&\underline{\mathcal E}^{\underline\alpha}=e^{-{1\over 6}\phi}\,{\mathcal
E}^{\underline\alpha}+e^{{1\over 6}\phi}\,\underline{\mathcal E}^{11}\,(\Gamma^{11}\lambda)^{\underline\alpha}
\,,\nonumber
\eee
where the index $11$ is identified with the index $7$ of the $U(1)$
fiber direction of $S^7$ and
\be\label{IIAA}
{{\mathcal A}}_1(x,y,\vartheta,\upsilon)=e^{-{2\over
3}\phi(\upsilon)}\,dZ^{\mathcal M}\,(E_{\mathcal
M}{}^a\,\Lambda_a{}^{7}+E_{\mathcal M}{}^7\,\Lambda_7{}^7)\,.
\ee
The one forms ${\mathcal E}^{\mathcal A}(x,y,\theta)=({\mathcal
E}^A,\,{\mathcal E}^{\underline\alpha})$, ${{\mathcal
A}}_1(x,y,\vartheta,\upsilon)$, the spinor superfield
$\lambda_{\underline\alpha}(x,y,\theta)$, with non-zero components
\be
\lambda_{\alpha i}=-\frac{1}{3}D_{\alpha i}\,\phi(\upsilon)\,,
\ee
and the scalar superfield $\phi(x,y,\theta)$ do not depend on the
11th coordinate $z$. They describe, respectively, the
supervielbeins, the RR one--form gauge superfield, the dilatino and
the dilaton superfields of type IIA supergravity in the \emph{string
frame}, eqs. (\ref{sva}) and (\ref{RR1form}). The RR field strength
$F_4$ and the NS--NS field strength $H_3$ given in eqs. (\ref{f4h3})
are obtained from the $D=11$ four--form field strength by the
conventional dimensional reduction described in \cite{Howe:2004ib}.
By construction they solve the type IIA supergravity constraints and
describe the $AdS_4\times CP^3$ background which preserves 24
supersymmetries. The explicit form of these and other relevant IIA
superfields has been given in Section
\ref{superadscp3}. Using this $AdS_4\times CP^3$ supergeometry, we
can write down the complete Green--Schwarz--type action for the
superstring and D--branes on this background (Section $3$).

\section{Conclusion}
We have   constructed the complete type IIA superspace with 32
fermionic coordinates which describes the  $AdS_4\times CP^3$ vacuum
solution of IIA supergravity preserving    24 supersymmetries in
terms of superfields depending on 32 fermionic coordinates. Our
construction guarantees that the geometry of this superspace and the
vacuum configurations of NS--NS and RR superfields living in it
solve the type IIA supergravity constraints (and therefore the full
set of type IIA equations of motion).\footnote{As an alternative
procedure of deriving this supergeometry one might try to directly
solve the type IIA supergravity constraints up to the 32-nd order in
fermionic variables taking the 24--component $OSp(6|4)/U(3)\times
SO(1,3)$ solution as the initial condition.} An important
qualitative difference with previous constructions of
supergeometries   is that the $AdS_4\times CP^3$ superspace is not a
coset space and that the type IIA $AdS_4\times CP^3$ superbackground
is not maximally supersymmetric.

Having the explicit form of the type IIA $AdS_4\times CP^3$
supergeometry has allowed us to write down the Green--Schwarz--type
action for the superstring and D--branes propagating in this
background. This provides  us with a concrete framework in which  to
study the most general  classical and quantum dynamics of these
branes. These actions complete  to the full 32--component superspace
the string sigma--model actions based on the $OSp(6|4)/U(3)\times
SO(1,3)$ supercoset constructed and studied in
\cite{Arutyunov:2008if,Stefanski:2008ik,Fre:2008qc,D'Auria:2008cw}.

We have analyzed the integrability of the classical equations of
motion of the superstring in different submanifolds of the full
$AdS_4\times CP^3$ superspace. For the  submanifold described by the
$OSp(6|4)/U(3)\times SO(1,3)$ supercoset, the classical equations of
motion are integrable,  as already has been  shown in
\cite{Arutyunov:2008if,Stefanski:2008ik}   following the
integrability criteria for sigma--models based on supercosets
discovered by Bena, Polchinski and Roiban \cite{Bena:2003wd}. We
have also  considered the supergeometry corresponding to the
``complementary" submanifold in   $AdS_4\times CP^3$ superspace.
Here we find that this sector of the theory is not based on a
supercoset, but on a ``twisted" $OSp(2|4)/SO(2)\times SO(1,3)$
superspace, whose supergeometry we have explicitly constructed by
restricting the total superspace to this submanifold. Whether the
equations of motion in this sector of the theory are classical
integrable remains    an important open problem. The fact that the
complete $AdS_4\times CP^3$ superspace is not a coset space requires
that more general methods are used to prove whether the superstring
equations of motion are classically integrable. The explicit
construction  in this paper of the geometry for the $AdS_4\times
CP^3$ superspace  provides a framework in which to study  this
problem.

Another important question for the future is to understand  whether
the classical dynamics of the string worldsheet can be encoded in
the Hamiltonian describing the spectrum of anomalous dimensions of
the holographic dual ABJM theory, extending to this holographic
correspondence the analogous results found for the $AdS_5/CFT_4$
correspondence. It also remains a challenge to find more arguments
in favour of the exact integrability of the planar dilatation
operator in the ABJM theory. The ultimate fate of the classical
integrability of the Green-Schwarz superstring action in
$AdS_4\times CP^3$ and the integrability of the planar ABJM
dilatation operator  are likely  to be related, and remain amongst the most important
open problems in this new holographic correspondence.

\section*{Acknowledgments}
J.G. would like to thank Joaquim Gomis and D.S. would like to thank
I. Bandos, M. Cederwall, P.A. Grassi, B. Nilsson, P. Pasti, M. Tonin
and M. Vasiliev for valuable discussions. D.S. is thankful to the
Perimeter Institute for kind hospitality and financial support of
his stay in which the main part of this work was accomplished.
Research at Perimeter Institute for Theoretical Physics is supported
in part by the Government of Canada through NSERC and by the
Province of Ontario through MRI. J.G. also acknowledges further
support from an NSERC Discovery Grant. Work of D.S. and L.W. was
partially supported by the INFN Special Initiative TV12, by the
European Commission FP6 program MRTN-CT-2004-005104 and by the INTAS
Project Grant 05-1000008-7928 in which D.S. and L.W. were associated
to the Department of Physics ``Galileo Galilei" of Padova
University. D.S. was also partially supported by an Excellence Grant
of Fondazione Cariparo and the grant FIS2008-1980 of the Spanish
Ministry of Science and Innovation.

\medskip
\medskip

\def\theequation{A.\arabic{equation}}
\def\thesection{Appendix A. Main notation and conventions}
\section{}
\setcounter{equation}0

The convention for the ten and eleven dimensional metrics is the
`almost plus' signature $(-,+,\cdots,+)$. Generically, the tangent
space vector indices are labeled by letters from the beginning of
the Latin alphabet,  while  letters from the middle of the Latin
alphabet stand for curved (world) indices. The spinor indices are
labeled by Greek letters.

\subsection*{$AdS_4$ space}

$AdS_4$ is parametrized by the coordinates $x^m$ and its vielbeins
are $e^a=dx^m\,e_m{}^a(x)$, $m=0,1,2,3;$ $a=0,1,2,3$. The $D=4$
gamma--matrices satisfy:
\be\label{gammaa}
\{\gamma^a,\gamma^b\}=2\,\eta^{ab}\,,\qquad \eta^{ab}={\rm
diag}\,(-,+,+,+)\,,
\ee
\be\label{gamma5}
\gamma^5=i\gamma^0\,\gamma^1\,\gamma^2\,\gamma^3, \qquad
\gamma^5\,\gamma^5=1\,.
\ee
The charge conjugation matrix $C$ is antisymmetric, the matrices
$(\gamma^a)_{\alpha\beta}\equiv (C\,\gamma^a)_{\alpha\beta}$ and
$(\gamma^{ab})_{\alpha\beta}\equiv(C\,\gamma^{ab})_{\alpha\beta}$
are symmetric and $\gamma^5_{\alpha\beta}\equiv
(C\gamma^5)_{\alpha\beta}$ is antisymmetric, with
$\alpha,\beta=1,2,3,4$ being the indices of a 4--dimensional spinor
representation of $SO(1,3)$ or $SO(2,3)$.

\subsection*{\bf $CP^3$ space}

$CP^3$ is parametrized by the coordinates $y^{m'}$ and its vielbeins
are $e^{a'}=dy^{m'}e_{m'}{}^{a'}(y)$, ${m'}=1,\cdots,6;$
${a'}=1,\cdots,6$. The $D=6$ gamma--matrices satisfy:
\be\label{gammaa'}
\{\gamma^{a'},\gamma^{b'}\}=2\,\delta^{{a'}{b'}}\,,\qquad \delta^{a'b'}={\rm
diag}\,(+,+,+,+,+,+)\,,
\ee
\be\label{gamma7}
\gamma^7={i\over{6!}}\,\epsilon_{\,a_1'a_2'a_3'a_4'a_5'a_6'}\,\gamma^{a_1'}\cdots \gamma^{a_6'} \qquad
\gamma^7\,\gamma^7=1\,.
\ee
The charge conjugation matrix $C'$ is symmetric and the matrices
$(\gamma^{a'})_{\alpha'\beta'}\equiv
(C\,\gamma^{a'})_{\alpha'\beta'}$ and
$(\gamma^{a'b'})_{\alpha'\beta'}\equiv(C'\,\gamma^{a'b'})_{\alpha'\beta'}$
are antisymmetric, with $\alpha',\beta'=1,\cdots,8$ being the
indices of an 8--dimensional spinor representation of $SO(6)$ or
$SO(8)$.

\subsection*{Seven--sphere}

$S^7$ is parametrized by the coordinates ${\hat y}^{{\hat
m}'}=(y^{m'},z)$, where $z$ stands for the coordinate of the Hopf
fiber in the description of $S^7$ as a $U(1)$ bundle over $CP^3$,
and its vielbeins are $e^{{\hat a}'}=d{\hat y}^{{\hat m}'}\,e_{{\hat
m}'}{}^{{\hat a}'}(\hat y)$, ${\hat m}'=(m',7);$ ${\hat a}'=(a',7)$.
The $D=7$ gamma--matrices are given by
\be\label{gammahata'}
\gamma^{{\hat a}'}=(\gamma^{a'},\gamma^7)\,,
\ee
and  satisfy the Clifford algebra
\be\label{gammaaa'}
\{\gamma^{{\hat a}'},\gamma^{{\hat b}'}\}=2\,\delta^{{{\hat a}'}{{\hat b}'}}\,,\qquad \delta^{{\hat a}'{\hat b}'}={\rm
diag}\,(+,+,+,+,+,+,+)\,.
\ee

\subsection*{\bf Type IIA  $AdS_4\times CP^3$ superspace}

The type IIA superspace whose bosonic body is $AdS_4\times CP^3$ is
parametrized by 10 bosonic coordinates $X^M=(x^m,\,y^{m'})$ and
32-fermionic coordinates $\theta^{\underline\mu}=(\theta^{\mu\mu'})$
($\mu=1,2,3,4;\,\mu'=1,\cdots,8$). These  combine into the
superspace supercoordinates $Z^{\cal
M}=(x^m,\,y^{m'},\,\theta^{\mu\mu'})$. The type IIA supervielbeins
are
\begin{equation}\label{IIAsv}
{\mathcal E}^{\mathcal A}=dZ^{\mathcal M}\,{\mathcal E}_{\mathcal M}{}^{\mathcal
A}(Z)=({\mathcal E}^{A},\,{\mathcal E}^{\underline\alpha})\,,\qquad
{\mathcal E}^{A}(Z)=({\mathcal E}^{a},\,{\mathcal E}^{a'})\,,\qquad
{\mathcal E}^{\underline\alpha}(Z)={\mathcal E}^{\alpha\alpha'}\,.
\ee
The $D=10$ gamma--matrices $\Gamma^A$ are given by
\bee\label{Gamma10}
&\{\Gamma^A,\,\Gamma^B\}=2\eta^{AB},\qquad
\Gamma^{A}=(\Gamma^a,\,\Gamma^{a'})\,,\nonumber\\
&\\
&\Gamma^{a}=\gamma^a\,\otimes\,{\bf 1},\qquad
\Gamma^{a'}=\gamma^5\,\otimes\,\gamma^{a'},\qquad
\Gamma^{11}=\gamma^5\,\otimes\,\gamma^7,\qquad a=0,1,2,3;\quad
a'=1,\cdots,6\,. \nonumber
\eee
The charge conjugation matrix is ${\mathcal C}=C\otimes C'$.

\subsection* {\bf Torsion constraint}

Our convention for the essential constraint on the torsion
$D\,{\mathcal E}^{\mathcal A}={1\over 2}\,{\mathcal E}^{\mathcal C}\,{\mathcal E}^{\mathcal
B}\,T_{\mathcal {BC}}{}^{\mathcal A}$ of IIA supergravity is
$T_{\underline\alpha\underline\beta}^A=2\Gamma_{\underline\alpha\underline\beta}^A$.
This choice is related to the form of the $OSp(8|4)$ algebra
(Appendix B, eq. (\ref{QQ})) and differs from that of
\cite{Howe:2004ib} by the factor $2i$.

\subsection*{Explicit form of the vielbeins and connections of
$OSp(6|4)/U(3)\times SO(1,3)$}

The Cartan form is
\bee
K^{-1}_{_{10,24}}\,dK_{_{10,24}} &=&
E_{_{10,24}}^a\,P_a+E_{_{10,24}}^{a'}\,P_{a'}+E_{_{10,24}}^{\alpha
a'}\,Q_{\alpha a'}
\nonumber\\
\\
&+&{1\over 2}\Omega_{_{10,24}}^{ab}\,M_{ab} +{1\over
2}\,\Omega_{_{10,24}}^{a'b'}\,(L_{a'b'}
-\frac{1}{6}\,J_{a'b'}\,J^{c'd'}L_{c'd'}) +A_{_{10,24}}\,T_1.
\nonumber
\eee
Computing these quantities explicitly using the commutation
relations (\ref{osp64}), the form of the $SU(4)$ generators of
Appendix C.2 and applying the method described
\emph{e.g.} in \cite{Kallosh:1998zx,Metsaev:1998hf,deWit:1998tk,Claus:1998fh} one finds
\be\label{cartan24}
\begin{aligned}
E_{_{10,24}}^a&=e^a(x)-4\vartheta\gamma^a\,{{\sinh^2{{\mathcal M}_{24}/ 2}}\over{\mathcal M}^2_{24}}\,
D_{24}\vartheta,\\
E_{_{10,24}}^{a'}&=e^{a'}(y)-4\vartheta\gamma^{a'}\gamma^5\,{{\sinh^2{{\mathcal
M}_{24}/2}}\over{\mathcal M}_{24}^2}\,D_{24}\vartheta\,,
\\
E_{_{10,24}}^{\alpha a'}&=\left({{\sinh{\mathcal
M}_{24}}\over{\mathcal
M}_{24}}D_{24}\vartheta\right)^{\alpha a'},\\
\Omega_{_{10,24}}^{ab}&=\omega^{ab}(x)+8i\vartheta\gamma^{ab}\gamma^5\,
{{\sinh^2{{\mathcal M}_{24}/2}}\over{\mathcal M}_{24}^2}D_{24}\vartheta\,,\\
\Omega_{_{10,24}}^{a'b'}&=\omega^{a'b'}(y)
-4i\vartheta(\gamma^{a'b'}-{i}J^{a'b'}\gamma^7)\gamma^5\,{{\sinh^2{{\mathcal
M}_{24}/2}}
\over{\mathcal M}_{24}^2}\,D_{24}\vartheta\,,\\
A_{_{10,24}}&={1\over
8}J_{a'b'}\Omega_{_{10,24}}^{a'b'}=A(y)-4\,\vartheta\gamma^7\gamma^5\,{{\sinh^2{{\mathcal
M}_{24}/2}}\over{\mathcal M}_{24}^2}\,D_{24}\vartheta\,,
\end{aligned}
\ee
where
\be
({\mathcal M}_{_{24}}^2)^{\alpha a'}{}_{\beta b'}=
4i\vartheta^{\alpha}_ {b'}\,(\vartheta^{a'}\gamma^5)_\beta
-4i\delta^{a'}_{b'}\vartheta^{\alpha c'}(\vartheta\gamma^5)_{\beta
c'} -2i(\gamma^5\gamma^a\vartheta)^{\alpha
a'}(\vartheta\gamma_a)_{\beta b'} -i(\gamma^{ab}\vartheta)^{\alpha
a'}(\vartheta\gamma_{ab}\gamma^5)_{\beta b'}\,.
\ee
The derivative appearing in the above equations is defined as
\be
D_{24}\vartheta={\mathcal P}_6\,(d +ie^a\gamma^5\gamma_a
+{i}e^{a'}\,\gamma_{a'}-\frac{1}{4}\omega^{ab}\gamma_{ab}-\frac{1}{4}\omega^{a'b'}
\,\gamma_{a'b'})\vartheta\,,
\ee
where $e^a(x)$, $e^{a'}(y)$, $\omega^{ab}(x)$, $\omega^{a'b'}(y)$
and $A(y)$ are the vielbeins and connections of the bosonic
$AdS_4\times CP^3$ solution (see Section \ref{s7fibration}).

The $U(3)$--connection
$\Omega_{_{10,24}}^{a'b'}=\Omega_{_{SU(3)}}^{a'b'}+\frac{4}{3}\,A_{_{10,24}}\,J^{a'b'}$
satisfies the condition
\begin{equation}
{(P^{-})_{a'b'}}^{c'd'}\Omega_{c'd'}=\frac{1}{2}\,({\delta_{[a'}}^{c'}\,{\delta_{b']}}^{d'}\,-\,
{J_{[a'}}^{c'}\,{J_{b']}}^{d'})\Omega_{c'd'}=0\,,
\end{equation}
where $J_{a'b'}$ is the K\"ahler form on $CP^3$.  Remember also that
$\vartheta={\mathcal P}_6\,\theta$ (see eqs. (\ref{p6}) and
(\ref{24})).

\subsection*{\bf Superspace  $OSp(8|4)/SO(7)\times SO(1,3)$}

Its bosonic body is $AdS_4\times S^7$ and it is parametrized by the
supercoordinates \linebreak ${\hat Z}^{\hat{\mathcal
M}}=({Z}^{\mathcal M},\,z)=(x^m,\,y^{m'},\,z,\,\theta^{\mu\mu'})$.
The corresponding supervielbeins are
\begin{equation}\label{11sv}
{E}_{_{11|32}}^{\hat{\mathcal A}}=d{\hat Z}^{\hat{\mathcal
M}}\,{\hat E}_{\hat{\mathcal M}}{}^{\hat{\mathcal A}}(\hat
Z)=d{Z}^{{\mathcal M}}\,{\hat E}_{{\mathcal M}}{}^{\hat{\mathcal
A}}(\hat Z)+dz\,{\hat E}_7{}^{\hat{\mathcal A}}(\hat Z)=({\hat
E}^{A},\,{\hat E}^{7},\,{\hat E}^{\alpha\alpha'})\,.
\ee
The label $7$ stands for the 7th direction along $S^7$ and 11-th
direction of $D=11$.
\\

\def\theequation{B.\arabic{equation}}
\def\thesection{Appendix B.~~$OSp(8|4)$, $OSp(2|4)$ and $OSp(6|4)$}
\section{}
\setcounter{equation}0
\subsection*{$OSp(8|4)$ superalgebra\footnote{Our conventions are similar to those in
\cite{Hatsuda:2002kx} modulo the minus sign in the definition of the generators of $SO(1,3)$ and
$SO(8)$.}}
This superalgebra consists of the following:
\\
\\
{\bf $SO(2,3)\simeq Sp(4)$ subalgebra}
\be\label{sp411}
[P_a,P_b]=-4M_{ab},\qquad [M_{ab},M_{cd}]=\eta_{ac}\,M_{bd}+
\eta_{bd}\,M_{ac}-\eta_{bc}\,M_{ad}-\eta_{ad}\,M_{bc}\,,
\ee
\be
[M_{ab},P_c]=\eta_{ac}\,P_b-\eta_{bc}\,P_a\,
\ee
where $P_a$ are the generators of  $AdS_4$  translations and
$M_{ab}$ are the generators of SO(1,3).

\subsection*{$SO(8)$ subalgebra}

\be\label{so7algebra1}
[ {  M}_{{\tilde a}'\,{\tilde b}'},\, {  M}_{{\tilde c}'\,{\tilde
d}'}] =\delta_{{\tilde a}'\,{\tilde c}'}\, {  M}_{{\tilde
b}'\,{\tilde d}'} -\delta_{{\tilde b}'\,{\tilde c}'}\, {
M}_{{\tilde a}'\,{\tilde d}'}+
\delta_{{\tilde b}'\,{\tilde d}'}\,  {  M}_{{\tilde a}'\,{\tilde c}'}- \delta_{{\tilde a}'\,{\tilde d}'}\, {  M}_{{\tilde b}'\,{\tilde c}'}\,.
\ee
where
\be\label{so81}
{  M}_{{\tilde a}'{\tilde
b}'}=(M_{a'b'},\,M_{a'7},\,M_{a'8},\,M_{78})\,,
\ee
and $M_{a'b'}$ $(a',b'=1,\cdots,6)$ are the generators of $SO(6)$.

\subsection*{Supersymmetry generators $Q_{\alpha\alpha'}$ in $OSp(8|4)$}
\be\label{PQ}
[P_a,\,Q_{\alpha\alpha'}]=i(Q_{\alpha'}\,\gamma^5\gamma_a)_\alpha\,,\qquad
[M_{ab}\,,Q_{\alpha\alpha'}]=-{1\over
2}\,(Q_{\alpha'}\,\gamma_{ab})_\alpha\,,
\ee
\be\label{MQ}
[{  M}_{{\tilde a}'{\tilde b}'}\,,Q_{\alpha\alpha'}]=-{1\over
2}\,(Q_{\alpha}\,\tilde\gamma_{{\tilde a}'{\tilde b}'})_{\alpha'}\,,
\ee
\be\label{QQ}
\{Q_{\alpha\alpha'},\,Q_{\beta\beta'}\}=-2\,C'_{\alpha'\beta'}\,(\gamma^a_{\alpha\beta}\,P_a
-i(\gamma^5\gamma^{ab})_{\alpha\beta}\,
M_{ab})-i\gamma^5_{\alpha\beta}\,(\tilde\gamma^{{\tilde a}'{\tilde
b}'})_{\alpha'\beta'}\,{  M}_{{\tilde a}'{\tilde b}'},
\ee
\medskip
where $\alpha=1,2,3,4$ are $Spin(2,3)$ indices and
$\alpha'=1,\cdots, 8$ are $Spin(8)$ indices. We remind the reader
that the matrices $C'_{\alpha'\beta'}$,
$\gamma^a_{\alpha\beta}=(C\gamma^a)_{\alpha\beta}$ and
$\gamma^{ab}_{\alpha\beta}\equiv(C\gamma^{ab})_{\alpha\beta}$ are
symmetric in spinor indices and the matrices $C_{\alpha\beta}$,
$\gamma^5_{\alpha\beta}\equiv(C\gamma^5)_{\alpha\beta}$ and
$(\tilde\gamma^{{\tilde a}'{\tilde b}'})_{\alpha'\beta'}$ are
antisymmetric. The $8\times 8$ matrices $\tilde\gamma^{{\tilde
a}'{\tilde b}'}$ -- which generate $SO(8)$ -- are given by
\be\label{tildeg}
\tilde\gamma^{{\tilde a}'{\tilde b}'}=-\tilde\gamma^{{\tilde b}'{\tilde a}'}
=(\gamma^{a'b'},\,\gamma^{a'7},\,\gamma^{a'8},\,\gamma^{78})\,,\qquad
\gamma^{a'8}\equiv {i}\,\gamma^{a'}\,,\qquad
\gamma^{78}\equiv {i}\,\gamma^7.
\ee

\subsection*{$OSp(2|4)$ superalgebra}

This algebra has 8 Grassmann--odd generators ${\mathcal Q}_{\alpha
i}$ $(i=1,2)$ which obey the following \linebreak (anti)commutation
relations
\be\label{PQ24}
[P_a,\,{\mathcal Q}_{\alpha i}]=i({\mathcal
Q}_{i}\,\gamma^5\gamma_a)_\alpha\,,\qquad [M_{ab},\,{\mathcal
Q}_{\alpha i}]=-{1\over 2}\,({\mathcal
Q}_{i}\,\gamma_{ab})_\alpha\,,
\ee
\be\label{TQ}
[T_2,{\mathcal Q}_{\alpha i}]=2\epsilon_i{}^{j}\,{\mathcal Q}_{\alpha
j}\,,
\ee
\be\label{QQ24}
\{{\mathcal Q}_{\alpha i},\,{\mathcal Q}_{\beta j}\}=
-2\,\delta_{ij}\,(\gamma^a_{\alpha\beta}\,P_a
-i(\gamma^5\gamma^{ab})_{\alpha\beta}\, M_{ab})- 2i
\gamma^5_{\alpha\beta}\,\epsilon_{ij}\,T_2,
\ee
where   $P_a$ and $M_{ab}$ are the generators of $SO(2,3)$  and
$T_2$ is the generator of $SO(2)$ and
$\epsilon_{ij}=-\epsilon_{ji}$, $\epsilon_{12}=1$ .

As a subalgebra of $OSp(8|4)$ the superalgebra $OSp(2|4)$ can be
obtained from eqs. (\ref{PQ})--(\ref{QQ}) by singling out 8
fermionic generators ${\mathcal Q_{\alpha i}}$ from the 32
generators $Q_{\alpha\alpha'}$ by applying to the latter the projector
${\mathcal P}_2$ which has two non--zero eigenvalues (see Appendix
C.2 for more details)
\be\label{p22}
{\mathcal P}_2={1\over 8}\,(2+J)\,,\qquad
J=-iJ_{a'b'}\,\gamma^{a'b'}\,\gamma^7\,,
\ee
\be\label{Q8}
({\mathcal P}_2\,Q)_{\alpha\alpha'}\qquad \Longleftrightarrow \qquad
{\mathcal Q_{\alpha i}}\,,
\ee
where $J_{a'b'}$ are components of the K\"ahler form on $CP^3$. Thus,
there is the following correspondence between the quantities
appearing in (\ref{PQ})--(\ref{tildeg}) and in
(\ref{PQ24})--(\ref{QQ24})
\be\label{co1}
T_2=-{1\over 2}\,(J^{a'b'}\,M_{a'b'}+2\,M_{78})\,,\qquad ({\mathcal
P}_2\,C'\,{\mathcal P}_2)_{\alpha'\beta'}\quad\Leftrightarrow
\quad \delta_{ij}
\,,\qquad ({\mathcal P}_2\,\gamma^7\,{\mathcal P}_2)_{\alpha'\beta'}
\quad\Leftrightarrow \quad
i\epsilon_{ij}\,.
\ee
\subsection*{$OSp(6|4)$ superalgebra}

This algebra has 24 Grassmann--odd generators ${Q}_{\alpha a'}$
$(a'=1,\cdots,6)$ which obey the following (anti)commutation
relations
\bee\label{osp64}
&[P_a,\,Q_{\alpha a'}]=i(Q_{a'}{\gamma^5\gamma_a})_\alpha\,,\quad
[M_{ab},\,Q_{\alpha a'}]=-{1\over
2}\,(Q_{a'}\,\gamma_{ab})_\alpha\,\,,\quad \nonumber\\
&\nonumber\\
&[M_{a'b'},\,Q_{\alpha c'}]=\,\delta_{a'c'}\,Q_{\alpha
b'}-\,\delta_{b'c'}\,Q_{\alpha a'}\,,\\
\nonumber\\
&\{Q_{\alpha a'},Q_{\beta
b'}\}=-2\,\delta_{a'b'}\,(\gamma^a_{\alpha\beta}\,P_a
-i(\gamma^5\gamma^{ab})_{\alpha\beta}\, M_{ab})
-4i\,\gamma^5_{\alpha\beta}\,M_{a'b'}\,,\nonumber
\eee
where $P_a$ and $M_{ab}$ are the generators of $SO(2,3)$   and
$M_{a'b'}$ are the generators of $SO(6)$
\begin{equation}\label{so6}
[M_{{a'}\,{b'}},\,M_{{c'}\,{d'}}]
=\delta_{{a'}\,{c'}}\,M_{{b'}\,{d'}} -\delta_{{b'}\,{c'}}\,
M_{{a'}\,{d'}}+
\delta_{{b'}\,{d'}}\, M_{{a'}\,{c'}}- \delta_{{a'}\,{d'}}\,M_{{b'}\,{c'}}\,.
\end{equation}

As a subalgebra of $OSp(8|4)$ the superalgebra $OSp(6|4)$ can be
obtained from eqs. (\ref{PQ})--(\ref{QQ}) by singling out 24
fermionic generators ${Q_{\alpha a'}}$ from the 32 generators
$Q_{\alpha\alpha'}$ by applying to the latter the projector
${\mathcal P}_6$ which has six non--zero eigenvalues (see Appendix
C.2 for more details)
\be\label{p62}
{\mathcal P}_6={1\over 8}\,(6-J)\,,\qquad
J=-iJ_{a'b'}\,\gamma^{a'b'}\,\gamma^7\,,
\ee
\be\label{Q24}
({\mathcal P}_6\,Q)_{\alpha\alpha'}\qquad \Longleftrightarrow \qquad
{ Q_{\alpha a'}}\,.
\ee
Thus, there is the following correspondence between the $SO(8)$
generators appearing in (\ref{PQ})--(\ref{tildeg}) and the $SO(6)$
generators appearing in (\ref{osp64})
\be\label{so61}
{1\over 4}\,({\mathcal P}_6\,\tilde\gamma^{{\tilde a}'{\tilde
b}'}\,{\mathcal P}_6)_{\alpha'\beta'}\,{M}_{{\tilde a}'{\tilde b}'}
\quad \Longleftrightarrow \quad M_{a'b'}\,,\qquad
({\mathcal P}_6\,C'\,{\mathcal P}_6)_{\alpha'\beta'}\quad\Longleftrightarrow
\quad \delta_{a'b'}.
\ee
In particular, the generator $T_1$ of the $U(1)$ subgroup of the
$CP^3$ structure group, which appeared in Sections 4 and 5, is
\be\label{T1}
T_1=\frac{1}{6}J^{a'b'}M_{a'b'}-M_{78}\,.
\ee

\subsection*{$OSp(8|4)$  closure of $Q_{\alpha a'}$ and ${\mathcal Q}_{\alpha i}$}

The anticommutator  of $Q_{\alpha a'}$ and ${\mathcal Q}_{\alpha i}$
\be\label{q6q2}
\{{Q_{\alpha a'}},\,{\mathcal Q}_{\beta
i}\}=-4i\,\gamma^5_{\alpha\beta}\,M_{a'i}\,
\ee
produces the generators
\be\label{ai}
M_{a'i}=({ M}_{a'7},\,{ M}_{a'8})\quad
\Longleftrightarrow \quad
{1\over{4}}\,({\mathcal P}_6\,\tilde\gamma^{{\tilde a}'{\tilde
b}'}\,{\mathcal P}_2)_{\alpha'\beta'}\,{M}_{{\tilde a}'{\tilde
b}'}\,
\ee
that correspond to the coset $SO(8)/SO(6)\times SO(2)$ and thus
complement the $SO(6)\times SO(2)$ generators $M_{a'b'}$ and $T_2$
(which can be associated with (redefined) ${M}_{78}$) to complete
the full $SO(8)$ algebra. Finally, the $OSp(2|4)$ and $OSp(6|4)$
superalgebras complete the full  $OSp(8|4)$ superalgebra with the
following commutation relations
\be\label{qmai}
[M_{a'i},\,Q_{\alpha b'}]=\delta_{a'b'}\,{\mathcal Q}_{\alpha
i}\,,\qquad [M_{a'i},\,{\mathcal Q}_{\alpha j}]=-\delta_{ij}\,{
Q}_{\alpha a'}\,.
\ee

\def\theequation{C.\arabic{equation}}
\def\thesection{Appendix C.~
$SU(3)\times U(1)$ embeddings into $SO(6)$}\label{B}
\section{}
\setcounter{equation}0

\def\thesection{C}
\subsection*{C.1 $SU(3)\times U(1)$ embedding into $SO(6)$ and the $CP^3$ coset
generators}\label{u3so6}

Let $M_{a'b'}=-M_{b'a'}$ ($a',b'=1,\cdots, 6$) be the 15 generators
of the $SO(6)$ algebra (\ref{so6}).

Let $J_{a'b'}=-J_{b'a'}$ be a constant antisymmetric matrix
(determining the components of the K\"ahler form on $CP^3$)
satisfying the relations
\begin{equation}\label{jmn}
J_{a'b'}=-J_{b'a'}\,,\qquad
J_{{a'}c'}\,{J^{c'}}_{b'}=-\delta_{{a'}{b'}}\,,\qquad
\epsilon_{{a'}{b'}c'd'e'f'}\,J^{{a'}{b'}}\,J^{c'd'}=8\,J_{e'f'}.
\end{equation}
Let $({P^{\pm})_{a'b'}}^{c'd'}$ be the following $15 \times 15$
projection matrices
\begin{equation}\label{ppm}
{(P^{\pm})_{a'b'}}^{c'd'}=\frac{1}{2}\,({\delta_{[a'}}^{c'}\,{\delta_{b']}}^{d'}\,\pm\,
{J_{[a'}}^{c'}\,{J_{b']}}^{d'}), \qquad P^++P^-=\mathbf{1}\,.
\end{equation}
The matrix $P^+$ has 9 non-zero eigenvalues and the matrix $P^-$ has
6 non-zero eigenvalues.

Then the generators
\be\label{u3}
L_{a'b'}={(P^{+})_{a'b'}}^{c'd'}\,M_{c'd'}
\ee
form the algebra $U(3)=SU(3)\times U(1)\subset SO(6)$ with $SU(3)$ generated by
\be\label{su3}
L_{a'b'}-\frac{1}{6}\,J_{a'b'}\,J^{c'd'}\,M_{c'd'}
\ee
and the $U(1)$ generated by
\be\label{u1}
T'=-\frac{1}{2}\,J^{c'd'}\,M_{c'd'}\,.
\ee
The remaining generators of $SU(4)\simeq Spin(6)$, namely
\be\label{cp3}
K_{a'b'}={(P^{-})_{a'b'}}^{c'd'}\,M_{c'd'}\,
\ee
form the coset space $CP^3=SU(4)/SU(3)\times U(1)$. They have the
following generic form of the commutation relations
\be\label{comK}
[K,K]=L,\qquad [K,L]=K\,.
\ee
For the construction of the $AdS_4 \times CP^3$ superspace we have,
however, used a different realization of the $SU(4)$ algebra
introduced below.

\subsection*{C.2 $SU(3)\times U(1)$ embedding into $Spin(6)$ and its extension to $SU(4)$  and $Spin(8)$ via $Spin(7)$.}
The necessity of understanding such an embedding is caused by the
fact that the 24 fermionic generators $Q$ of the $OSp(6|4)$
superalgebra (which is the super--isometry of the $AdS_4\times CP^3$
solution of IIA supergravity preserving 24 supersymmetries) have a
natural realization as a direct product of 4-dimensional spinors of
$Sp(4)\simeq Spin(2,3)$ and 6--dimensional vectors of $SO(6)$,
\emph{i.e.} $Q_{\alpha a'}$ carry the $Spin(2,3)$ spinor indices
$\alpha=1,2,3,4$ and $SO(6)$ vector indices $a'=1,\cdots, 6$. The
structure of the $OSp(6|4)$ superalgebra is given in eqs.
(\ref{osp64}).

At the same time the fermionic variables $\theta^{\underline\alpha}$
of IIA supergravity carry 32--component spinor indices of
$Spin(1,9)$ which in the $AdS_4\times CP^3$ background naturally
split into 4--dimensional $Spin(1,3)$ indices and 8--dimensional
spinor indices of $Spin(6)$, i.e.
$\theta^{\underline\alpha}=\theta^{\alpha\alpha'}$
($\alpha=1,2,3,4$; $\alpha'=1,\cdots,8$). 24 of these $\theta$'s
should correspond to the unbroken supersymmetries of the
$AdS_4\times CP^3$ background generated by the 24 $Q_{\alpha a'}$.

These 24 $\theta$ are singled out by a projector introduced in
\cite{Nilsson:1984bj} which is constructed using the K\"ahler
form (\ref{jmn}) and seven $8\times 8$ antisymmetric gamma--matrices
(\ref{gammaa'}). The $8\times 8$ projector matrix has the following
form
\be\label{p6}
{\mathcal P}_{6}={1\over 8}(6-J)\,,
\ee
where the $8\times 8$  matrix
\be\label{J}
J=-iJ_{a'b'}\,\gamma^{a'b'}\,\gamma^7 \qquad {\rm such~ that} \qquad
J^2= 4J+12
\ee
has six eigenvalues $-2$ and two eigenvalues $6$, \emph{i.e.} its
diagonalization results in
\be\label{Jdia}
J=\hbox{diag}(-2,-2,-2,-2,-2,-2,6,6)\,.
\ee
Therefore, the projector (\ref{p6}) when acting on an 8--dimensional
spinor annihilates 2 and leaves 6 of its components, while the
complementary projector
\be\label{p2}
{\mathcal P}_{2}={1\over 8}(2+J)\,,\qquad \mathcal{P}_2+\mathcal{P}_6=\mathbf 1
\ee
annihilates 6 and leaves 2 spinor components.

Thus the spinor
\be\label{24}
\vartheta^{\alpha\alpha'}=({\mathcal P}_6\,\theta)^{\alpha\alpha'} \qquad \Longleftrightarrow \qquad
\vartheta^{\alpha a'}\, \qquad a'=1,\cdots, 6
\ee
has 24 non--zero components and the spinor
\be\label{8}
\upsilon^{\alpha\alpha'}=({\mathcal P}_2\,\theta)^{\alpha\alpha'}\qquad \Longleftrightarrow \qquad
\upsilon^{\alpha i}\, \qquad i=1,2
\ee
has 8 non--zero components. The latter corresponds to the eight
supersymmetries broken by the $AdS_4\times CP^3$ background.

We would like to relate the 24--component fermionic variable
$\vartheta^{\alpha a'}$  to the
Grassmann--odd generators $Q_{\alpha a'}$ taking values in the
6--dimensional vector representation of $Spin(6)\simeq SU(4)$. To this
end, remember that the original fermionic variable
$\theta^{\alpha\alpha'}$ takes values in the 8--dimensional spinor
representation of $Spin(6)\simeq SU(4)$, generated by the antisymmetric
product of 6 gamma--matrices $\gamma^{a'}$
\be\label{mgmn}
M_{a'b'}=-\frac{1}{2}\,\gamma_{a'b'}\,,\qquad \gamma_{a'b'}\equiv
\frac{1}{2}\,(\gamma_{a'}\,\gamma_{b'}-\gamma_{b'}\,\gamma_{a'})\,.
\ee
The projected spinor (\ref{24}) will therefore transform by the
generators of the form
\be\label{p6ggp6}
L_{a'b'}=-\frac{1}{2}\,{\mathcal P}_6\,\gamma_{a'b'}\,{\mathcal
P}_6\,.
\ee
The question is what algebra is generated by (\ref{p6ggp6})?
Naively, one might think that it is again $Spin(6)\sim SU(4)$.
However, it turns out that only the generators of the $U(3)$
subgroup of $Spin(6)$ survive under the action of the projector
${\mathcal P}_6$. Namely, using the (anti)commutation relation of
$J$ (defined in (\ref{J})) with $\gamma^{a'}$
\be\label{Jg}
J\,\gamma^{a'}+\gamma^{a'}\,J=-4i\,J^{a'}{}_{b'}\,\gamma^{b'}\,\gamma^7\,,\qquad
[\gamma_{a'b'},\,J]=8i\,J_{[a'}{}^{c'}\,\gamma_{b']c'}\,\gamma^7
\ee
one can show that the following identities hold
\be\label{u3g}
L_{a'b'}=-\frac{1}{2}\,{\mathcal P}_6\,\gamma_{a'b'}\,{\mathcal
P}_6=-\frac{1}{2}\, {(P^{+})_{a'b'}}^{c'd'}\,{\mathcal
P}_6\,\gamma_{c'd'}\,{\mathcal P}_6,\,\quad
{(P^{-})_{a'b'}}^{c'd'}\,{\mathcal P}_6\,\gamma_{c'd'}\,{\mathcal
P}_6=0\,,
\ee
\be\label{p6ggp2}
\,{\mathcal P}_6\,\gamma_{a'b'}\,{\mathcal P}_2
={(P^{-})_{a'b'}}^{c'd'}\,{\mathcal P}_6\,\gamma_{c'd'}\,{\mathcal
P}_2\,,\qquad {(P^{+})_{a'b'}}^{c'd'}\,{\mathcal
P}_6\,\gamma_{c'd'}\,{\mathcal P}_2=0\,,
\ee
where $P^{\pm}$ were defined in (\ref{ppm}). Thus, in view of the
consideration of Subsection \ref{u3so6} the operators (\ref{p6ggp6})
indeed generate the $U(3)$ algebra, their $SU(3)$ and $U(1)$
subalgebras being generated, respectively, by\footnote{Note that in
the main text, for brevity, the $SU(3)$ generators associated with
(\ref{su31}) are denoted by $L_I$ (see \emph{e.g.} eqs.
(\ref{su4})--(\ref{cfs7}), (\ref{c1024})--(\ref{c11241}) and
(\ref{car1132})).}
\be\label{su31}
C_{a'b'}{}^I\,L_I= 2L_{a'b'}-{i\over {3}}\,J_{a'b'}\,{\mathcal
P}_6\,\gamma^7\,{\mathcal P}_6
\ee
and
\be\label{u11}
T'={1\over {4}}\,J_{a'b'}\,{\mathcal P}_6\,\gamma^{a'b'}\,{\mathcal
P}_6 =-\frac{i}{2}\,{\mathcal P}_6\,\gamma^7{\mathcal P}_6
\ee
(compare eqs. (\ref{su31}) and (\ref{u11}) with (\ref{su3}) and
(\ref{u1})).

Note that the $CP^3$ coset space generators (\ref{cp3}) do not
survive under the ${\mathcal P}_6$ projection. We should therefore
find another way to extend the $U(3)$ generators (\ref{p6ggp6}) to
$Spin(6)\simeq SU(4)$. It turns out that the matrices ${\mathcal
P}_6\,\gamma_{a'}\,\gamma^7 {\mathcal P}_6$ do this job, i.e. they
correspond to the six generators of the coset space
$CP^3=SU(4)/U(3)$. Indeed, using the identities
\be\label{p6gp6}
{\tilde P}_{a'}=-{\mathcal P}_6\,\gamma_{a'}\,\gamma^7\, {\mathcal
P}_6=-\frac{1}{2}\,(\delta_{a'}{}^{b'}-i\,J_{a'}{}^{b'}\,\gamma^7){\mathcal
P}_6\,\gamma_{b'}\,\gamma^7\, {\mathcal P}_6,
\ee
\be\label{p6gp2}
{\mathcal P}_6\,\gamma_{a'}\, {\mathcal
P}_2=\frac{1}{2}\,(\delta_{a'}{}^{b'}+i\,J_{a'}{}^{b'}\,\gamma^7){\mathcal
P}_6\,\gamma_{b'}\, {\mathcal P}_2\,,\quad {\mathcal
P}_2\,\gamma_{a'}\, {\mathcal
P}_6=\frac{1}{2}\,(\delta_{a'}{}^{b'}+i\,J_{a'}{}^{b'}\,\gamma^7){\mathcal
P}_2\,\gamma_{b'}\, {\mathcal P}_6
\ee
and
\be\label{p2gp2}
{\mathcal P}_2\,\gamma_{a'}\,{\mathcal P}_2=0
\ee
one can show that ${\tilde P}_{a'}$, defined in (\ref{p6gp6}), and
the $U(3)$ generators $L_{a'b'}$, defined in eq.(\ref{p6ggp6}), form
the following realization of the $Spin(6)\simeq SU(4)$ algebra
\be\label{u31}
[{\tilde P}_{a'},{\tilde P}_{b'}]=2L_{{a'}b'}, \qquad [{\tilde
P}_{a'},L_{b'c'}]=(\delta_{a'b'}-i\,J_{a'b'}\gamma^7)\,{\tilde
P}_{c'}-(\delta_{a'c'}-i\,J_{a'c'}\gamma^7)\,{\tilde P}_{b'}\,.
\ee
Note that instead of the generators ${\tilde P}_{a'}$ defined in
(\ref{p6gp6}) one can equivalently use the generators
\be\label{tildep}
P_{a'}=J_{a'}{}^{b'}\,{\tilde P}_{b'}={i}\,{\mathcal
P}_6\,\gamma_{a'} {\mathcal P}_6
\ee
as the $CP^3$ translations, as we actually do in the main part of
the paper.

The six generators $-\frac{1}{2}\gamma_{a'}\gamma_7$ extend
$Spin(6)\simeq  SU(4)$ to $Spin(7)$
\be\label{so7}
M_{{\hat a}'{\hat b}'}=(M_{a'b'},M_{a'7}), \qquad
M_{a'7}=-M_{7a'}=-{1\over 2}\,\gamma_{a'}\gamma_7, \qquad {\hat
a}'=({a'},7)
\ee
\be\label{so7algebra}
[ M_{{\hat a}'\,{\hat b}'},\, M_{{\hat c}'\,{\hat d}'}]
=\delta_{{\hat a}'\,{\hat c}'}\, M_{{\hat b}'\,{\hat d}'}
-\delta_{{\hat b}'\,{\hat c}'}\, M_{{\hat a}'\,{\hat d}'}+
\delta_{{\hat b}'\,{\hat d}'}\,  M_{{\hat a}'\,{\hat c}'}- \delta_{{\hat a}'\,{\hat d}'}\, M_{{\hat b}'\,{\hat c}'}\,.
\ee
Note also that the following matrices further extend the $Spin(7)$
algebra (\ref{so7}) to $Spin(8)$
\be\label{s7}
M_{a'8}=-M_{8a'}\equiv- {i\over 2}\,\gamma_{a'}\,,\qquad
M_{78}\equiv -{i\over 2}\gamma_7\,.
\ee
Namely, the $Spin(8)$ algebra is generated by
\be\label{so8}
{\mathcal M}_{{\tilde a}'{\tilde
b}'}=(M_{a'b'},\,M_{a'7},\,M_{a'8},\,M_{78})\,,
\ee
where $M_{a'8}$ and $M_{78}$, defined in (\ref{s7}), correspond to
an $S^7$--sphere coset $SO(8)/SO(7)$.

In terms of the generators $M_{a'7}$ and $M_{a'8}$, the $CP^3$
generators (\ref{p6gp6}) or (\ref{tildep}) are given by
$$
\tilde P_{a'}=M_{a'7}+J_{a'}{}^{b'}\,M_{b'8}, \qquad
 P_{a'}=-M_{a'8}+J_{a'}{}^{b'}\,M_{b'7}.
$$

Thus, to reduce 8--component spinors to 6--component ``vectors"
taking values in the corresponding representation of $Spin(6)\simeq
SU(4)$ one should start with the 8--component spinor representations
of the $Spin(7)$ algebra (\ref{so7}) and apply to them the projector
${\mathcal P}_6$ (\ref{p6}).

What about the ${\mathcal P}_2$ projection of $\gamma_{a'b'}$? It
has the form similar to eq. (\ref{u3g})
\be\label{p2ggp2}
\frac{1}{2}\,{\mathcal P}_2\,\gamma_{a'b'}\,{\mathcal P}_2
=\frac{1}{2}\,{(P^{+})_{a'b'}}^{c'd'}\,{\mathcal
P}_2\,\gamma_{c'd'}\,{\mathcal P}_2,
\ee
but now one should remember that ${\mathcal P}_2$ has only 2
non--zero eigenvalues and, hence, the matrix ${\mathcal
P}_2\,\gamma_{a'b'}\,{\mathcal P}_2$ is effectively a $2\times 2$
\emph{antisymmetric} matrix (in spinor indices). Since there is
only one independent $2\times 2$ antisymmetric matrix, the matrices
(\ref{p2ggp2}) belong to an $SO(2)\simeq U(1)$ algebra which commutes
with the $SU(4)\simeq SO(6)$ algebra generated by (\ref{p6ggp6}) and
(\ref{p6gp6}).

Thus, the generic form of the matrix (\ref{p2ggp2}) is
$X_{a'b'}\,\epsilon_{ij}$, where $X_{a'b'}$ and $\epsilon_{ij}$ is an
antisymmetric $6\times 6$ and $2\times 2$ matrix, respectively.
Since the only $U(3)$--invariant antisymmetric $6\times 6$ matrix is
$J_{a'b'}$, the matrices (\ref{p2ggp2}) actually reduce to
\be\label{p2ggp2J}
-\frac{1}{2}\,{\mathcal P}_2\,\gamma_{a'b'}\,{\mathcal P}_2=
-\frac{i}{12}\,J_{a'b'}
\,({\mathcal P}_2\,J\gamma^7\,{\mathcal P}_2)=-\frac{i}{2}\,
J_{a'b'}\,({\mathcal P}_2\,\gamma^7\,{\mathcal P}_2)\,,
\ee
which can also be checked directly using an explicit form of the
$\gamma^{a'}$--matrices. The Abelian algebra generated by the $2
\times 2$ antisymmetric matrix $-\frac{i}{2}\,\,{\mathcal
P}_2\,\gamma^7\,{\mathcal P}_2$ can be associated with the $SO(2)$
subalgebra of $SO(8)$ which commutes with $SO(6)$ generated by eq.
(\ref{u31}).


\begin{thebibliography}{999}
\bibitem{Bagger:2006sk}
J.~Bagger and N.~Lambert,
``Modeling multiple M2's,''
Phys.\ Rev.\  D {\bf 75} (2007) 045020
[arXiv:hep-th/0611108];
\\
J.~Bagger and N.~Lambert,
``Gauge Symmetry and Supersymmetry of Multiple M2-Branes,''
Phys.\ Rev.\  D {\bf 77} (2008) 065008
[arXiv:0711.0955 [hep-th]];
\\
J.~Bagger and N.~Lambert,
``Comments On Multiple M2-branes,''
JHEP {\bf 0802} (2008) 105
[arXiv:0712.3738 [hep-th]].
\bibitem{Gustavsson:2007vu}
A.~Gustavsson, ``Algebraic structures on parallel M2-branes,''
arXiv:0709.1260 [hep-th].
\bibitem{Gomis:2008uv}
J.~Gomis, G.~Milanesi and J.~G.~Russo,
``Bagger-Lambert Theory for General Lie Algebras,''
JHEP {\bf 0806} (2008) 075
[arXiv:0805.1012 [hep-th]];
\\
S.~Benvenuti, D.~Rodriguez-Gomez, E.~Tonni and H.~Verlinde,
``N=8 superconformal gauge theories and M2 branes,''
arXiv:0805.1087 [hep-th];
\\
P.~M.~Ho, Y.~Imamura and Y.~Matsuo,
``M2 to D2 revisited,''
JHEP {\bf 0807}, 003 (2008)
[arXiv:0805.1202 [hep-th]];
\\
M.~A.~Bandres, A.~E.~Lipstein and J.~H.~Schwarz,
``Ghost-Free Superconformal Action for Multiple M2-Branes,''
JHEP {\bf 0807} (2008) 117
[arXiv:0806.0054 [hep-th]];
\\
J.~Gomis, D.~Rodriguez-Gomez, M.~Van Raamsdonk and H.~Verlinde,
``Supersymmetric Yang-Mills Theory From Lorentzian Three-Algebras,''
JHEP {\bf 0808} (2008) 094
[arXiv:0806.0738 [hep-th]];
\\
B.~Ezhuthachan, S.~Mukhi and C.~Papageorgakis,
``D2 to D2,''
JHEP {\bf 0807}, 041 (2008)
[arXiv:0806.1639 [hep-th]];
\\
H.~Verlinde,
``D2 or M2? A Note on Membrane Scattering,''
arXiv:0807.2121 [hep-th].

\bibitem{Aharony:2008ug}
O.~Aharony, O.~Bergman, D.~L.~Jafferis and J.~Maldacena, ``N=6
superconformal Chern-Simons-matter theories, M2-branes and their
gravity duals,'' JHEP {\bf 0810} (2008) 091
  [arXiv:0806.1218 [hep-th]].


\bibitem{Gaiotto:2008sd}
D.~Gaiotto and E.~Witten,
``Janus Configurations, Chern-Simons Couplings, And The Theta-Angle in N=4
Super Yang-Mills Theory,''
arXiv:0804.2907 [hep-th];
\\
K.~Hosomichi, K.~M.~Lee, S.~Lee, S.~Lee and J.~Park,
``N=4 Superconformal Chern-Simons Theories with Hyper and Twisted Hyper
Multiplets,''
JHEP {\bf 0807}, 091 (2008)
[arXiv:0805.3662 [hep-th]].

\bibitem{Drukker:2008zx}
N.~Drukker, J.~Plefka and D.~Young,
``Wilson loops in 3-dimensional N=6 supersymmetric Chern-Simons Theory and
their string theory duals,''
arXiv:0809.2787 [hep-th].
\\
For Wilson loops in ABJM see also:
\\
 B.~Chen and J.~B.~Wu,
 ``Supersymmetric Wilson Loops in N=6 Super Chern-Simons-matter theory,''
 arXiv:0809.2863 [hep-th];
 \\
 S.~J.~Rey, T.~Suyama and S.~Yamaguchi,
 ``Wilson Loops in Superconformal Chern-Simons Theory and Fundamental Strings
 in Anti-de Sitter Supergravity Dual,''
 arXiv:0809.3786 [hep-th].

\bibitem{Drukker:2008jm}
N.~Drukker, J.~Gomis and D.~Young,
``Vortex Loop Operators, M2-branes and Holography,''
arXiv:0810.4344 [hep-th].

\bibitem{Nishioka:2008ib}
T.~Nishioka and T.~Takayanagi,
``Fuzzy Ring from M2-brane Giant Torus,''
JHEP {\bf 0810}, 082 (2008)
[arXiv:0808.2691 [hep-th]].
\bibitem{Berenstein:2008dc}
D.~Berenstein and D.~Trancanelli,
``Three-dimensional N=6 SCFT's and their membrane dynamics,''
arXiv:0808.2503 [hep-th].
\bibitem{Kluson:2008wn}
J.~Kluson and K.~L.~Panigrahi,
``Defects and Wilson Loops in 3d QFT from D-branes in AdS(4) x CP**3,''
arXiv:0809.3355 [hep-th].


\bibitem{Green:1983wt}
M.~B.~Green and J.~H.~Schwarz,
``Covariant Description Of Superstrings,''
Phys.\ Lett.\  B {\bf 136} (1984) 367;
\\
M.~B.~Green and J.~H.~Schwarz,
``Properties Of The Covariant Formulation Of Superstring Theories,''
Nucl.\ Phys.\  B {\bf 243} (1984) 285.
\bibitem{Grisaru:1985fv}
M.~T.~Grisaru, P.~S.~Howe, L.~Mezincescu, B.~Nilsson and P.~K.~Townsend,
``N=2 Superstrings in a Supergravity Background,''
Phys.\ Lett.\  B {\bf 162} (1985) 116.

\bibitem{Bergshoeff:1987cm}
E.~Bergshoeff, E.~Sezgin and P.~K.~Townsend,
``Supermembranes and eleven-dimensional supergravity,''
Phys.\ Lett.\  B {\bf 189} (1987) 75;
\\
E.~Bergshoeff, E.~Sezgin and P.~K.~Townsend,
``Properties of the Eleven-Dimensional Super Membrane Theory,''
Annals Phys.\  {\bf 185} (1988) 330.

\bibitem{Cederwall:1996pv}
M.~Cederwall, A.~von Gussich, B.~E.~W.~Nilsson and A.~Westerberg,
``The Dirichlet super-three-brane in ten-dimensional type IIB
supergravity,''
Nucl.\ Phys.\ B {\bf 490} (1997) 163
[arXiv:hep-th/9610148];
\\
M.~Cederwall, A.~von Gussich, B.~E.~W.~Nilsson, P.~Sundell and A.~Westerberg,
``The Dirichlet super-p-branes in ten-dimensional type IIA and IIB
supergravity,''
%
Nucl.\ Phys.\ B {\bf 490} (1997) 179
[arXiv:hep-th/9611159].

\bibitem{Aganagic:1996pe}
M.~Aganagic, C.~Popescu and J.~H.~Schwarz,
``D-brane actions with local kappa symmetry,''
%
Phys.\ Lett.\ B {\bf 393} (1997) 311
[arXiv:hep-th/9610249].


\bibitem{Bergshoeff:1996tu}
E.~Bergshoeff and P.~K.~Townsend,
``Super D-branes,''
%
Nucl.\ Phys.\ B {\bf 490} (1997) 145
[arXiv:hep-th/9611173].

\bibitem{Bandos:1997rq}
I.~A.~Bandos, D.~P.~Sorokin and M.~Tonin,
``Generalized action principle and superfield equations of motion for  D = 10
D p-branes,''
Nucl.\ Phys.\ B {\bf 497} (1997) 275
[arXiv:hep-th/9701127].

\bibitem{Bandos:1997ui}
I.~A.~Bandos, K.~Lechner, A.~Nurmagambetov, P.~Pasti, D.~P.~Sorokin and M.~Tonin,
``Covariant action for the super-five-brane of M-theory,''
%
Phys.\ Rev.\ Lett.\  {\bf 78} (1997) 4332
[arXiv:hep-th/9701149];
\\
M.~Aganagic, J.~Park, C.~Popescu and J.~H.~Schwarz,
``World-volume action of the M-theory five-brane,''
%
Nucl.\ Phys.\ B {\bf 496} (1997) 191
[arXiv:hep-th/9701166].

\bibitem{Cvetic:1999zs}
M.~Cvetic, H.~Lu, C.~N.~Pope and K.~S.~Stelle,
``T-duality in the Green-Schwarz formalism, and the massless/massive IIA
duality map,''
Nucl.\ Phys.\  B {\bf 573}, 149 (2000)
[arXiv:hep-th/9907202].

\bibitem{Metsaev:1998it}
R.~R.~Metsaev and A.~A.~Tseytlin,
``Type IIB superstring action in AdS(5) x S(5) background,''
Nucl.\ Phys.\  B {\bf 533} (1998) 109
[arXiv:hep-th/9805028].
\bibitem{Kallosh:1998zx}
R.~Kallosh, J.~Rahmfeld and A.~Rajaraman,
``Near horizon superspace,''
JHEP {\bf 9809} (1998) 002
[arXiv:hep-th/9805217].

\bibitem{Metsaev:1998hf}
R.~R.~Metsaev and A.~A.~Tseytlin,
``Supersymmetric D3 brane action in AdS(5) x S**5,''
Phys.\ Lett.\  B {\bf 436} (1998) 281
[arXiv:hep-th/9806095].

\bibitem{deWit:1998tk}
B.~de Wit, K.~Peeters and J.~Plefka,
``Superspace geometry for supermembrane backgrounds,''
Nucl.\ Phys.\  B {\bf 532} (1998) 99
[arXiv:hep-th/9803209];
\\
B.~de Wit, K.~Peeters, J.~Plefka and A.~Sevrin,
``The M-theory two-brane in AdS(4) x S(7) and AdS(7) x S(4),''
Phys.\ Lett.\  B {\bf 443} (1998) 153
[arXiv:hep-th/9808052].

\bibitem{Claus:1998fh}
P.~Claus,
``Super M-brane actions in AdS(4) x S(7) and AdS(7) x S(4),''
Phys.\ Rev.\  D {\bf 59} (1999) 066003
[arXiv:hep-th/9809045].


\bibitem{Pasti:1998tc}
P.~Pasti, D.~P.~Sorokin and M.~Tonin,
``On gauge-fixed superbrane actions in AdS superbackgrounds,''
Phys.\ Lett.\  B {\bf 447} (1999) 251
[arXiv:hep-th/9809213].

\bibitem{Arutyunov:2008if}
G.~Arutyunov and S.~Frolov,
``Superstrings on $AdS_4 \times CP^3$ as a Coset Sigma-model,''
JHEP {\bf 0809} (2008) 129
[arXiv:0806.4940 [hep-th]].

\bibitem{Stefanski:2008ik}
B.~j.~Stefanski,
``Green-Schwarz action for Type IIA strings on $AdS_4\times CP^3$,''
arXiv:0806.4948 [hep-th].

\bibitem{Fre:2008qc}
P.~Fr\'e and P.~A.~Grassi, ``Pure Spinor Formalism for ${Osp}(N|4)$
backgrounds,'' arXiv:0807.0044 [hep-th].

\bibitem{Bonelli:2008us}
 G.~Bonelli, P.~A.~Grassi and H.~Safaai,
 JHEP {\bf 0810}, 085 (2008)
 [arXiv:0808.1051 [hep-th]].

\bibitem{D'Auria:2008cw}
R.~D'Auria, P.~Fr\'e, P.~A.~Grassi and M.~Trigiante, ``Superstrings
on $AdS_4 x CP^3$ from Supergravity,'' arXiv:0808.1282 [hep-th].


\bibitem{Bena:2003wd}
I.~Bena, J.~Polchinski and R.~Roiban,
``Hidden symmetries of the AdS(5) x S**5 superstring,''
Phys.\ Rev.\  D {\bf 69}, 046002 (2004)
[arXiv:hep-th/0305116].



\bibitem{Watamura:1983ht}
S.~Watamura,
``Spontaneous Compactification of D = 10 Maxwell-Einstein Theory Leads to
SU(3) X SU(2) X U(1) Gauge Symmetry,''
Phys.\ Lett.\  B {\bf 129} (1983) 188.

\bibitem{Volkov:1984yw}
D.~V.~Volkov, D.~P.~Sorokin and V.~I.~Tkach,
``Mechanisms of Spontaneous Compactification of N=2, D = 10
Supergravitation,''
JETP Lett.\  {\bf 38}, 481 (1983)
[Pisma Zh.\ Eksp.\ Teor.\ Fiz.\  {\bf 38}, 397 (1983)].
\\
D.~V.~Volkov, D.~P.~Sorokin and V.~I.~Tkach,
``Spontaneous Compactification into Symmetric Spaces with Nonsimple Holonomy
Group,''
Theor.\ Math.\ Phys.\  {\bf 61} (1985) 1117
[Teor.\ Mat.\ Fiz.\  {\bf 61} (1984) 241].
\\
D.~V.~Volkov, D.~P.~Sorokin and V.~I.~Tkach,
``Spontaneous Compactification of Subspaces in Supergravity with D =
10, D = 11,''
Sov.\ J.\ Nucl.\ Phys.\  {\bf 39} (1984) 823
[Yad.\ Fiz.\  {\bf 39} (1984) 1306].


\bibitem{Campbell:1984zc}
I.~C.~G.~Campbell and P.~C.~West,
``N=2 D=10 Nonchiral Supergravity and its Spontaneous Compactification,''
Nucl.\ Phys.\  B {\bf 243} (1984) 112.

\bibitem{Watamura:1983hj}
S.~Watamura,
``Spontaneous Compactification and Cp(N): $SU(3)\times SU(2) \times U(1)$,
$sin^2\theta_W$, g(3)/g(2) and SU(3) Triplet Chiral Fermions in
Four-Dimensions,''
Phys.\ Lett.\  B {\bf 136} (1984) 245.


\bibitem{Freund:1980xh}
P.~G.~O.~Freund and M.~A.~Rubin,
``Dynamics of Dimensional Reduction,''
Phys.\ Lett.\  B {\bf 97}, 233 (1980).

\bibitem{Volkov:1980kq}
L.~V.~Volkov and V.~I.~Tkach,
``Spontaneous compactification of subspace due to interaction of the
Einstein fields with the gauge fields,''
JETP Lett.\  {\bf 32}, 668 (1980)
[Pisma Zh.\ Eksp.\ Teor.\ Fiz.\  {\bf 32}, 681 (1980)];
\\
D.~V.~Volkov and V.~I.~Tkach, ``Spontaneous Compactification of
Subspaces,'' Theor.\ Math.\ Phys.\  {\bf 51}, 427 (1982) [Teor.\
Mat.\ Fiz.\  {\bf 51}, 171 (1982)].

\bibitem{Giani:1984wc}
F.~Giani and M.~Pernici,
``N=2 Supergravity in Ten-Dimensions,''
Phys.\ Rev.\  D {\bf 30} (1984) 325.

\bibitem{Nilsson:1984bj}
B.~E.~W.~Nilsson and C.~N.~Pope,
``Hopf Fibration of Eleven-Dimensional Supergravity,''
Class.\ Quant.\ Grav.\  {\bf 1} (1984) 499.

\bibitem{Sorokin:1985ap}
 D.~P.~Sorokin, V.~I.~Tkach and D.~V.~Volkov,
 ``Kaluza-Klein Theories And Spontaneous Compactification Mechanisms Of Extra
 Space Dimensions,''
{\it  In *Moscow 1984, Proceedings, Quantum Gravity*, 376-392};
\\
D.~P.~Sorokin, V.~I.~Tkach and D.~V.~Volkov, ``On the Relationship
between Compactified Vacua of D = 11 and D = 10 Supergravities,''
Phys.\ Lett.\  B {\bf 161}, 301 (1985).


\bibitem{Aldazabal:2007sn}
G.~Aldazabal and A.~Font,
``A second look at N=1 supersymmetric $AdS_4$ vacua of type IIA supergravity,''
JHEP {\bf 0802} (2008) 086
[arXiv:0712.1021 [hep-th]].

\bibitem{Tomasiello:2007eq}
A.~Tomasiello,
``New string vacua from twistor spaces,''
Phys.\ Rev.\  D {\bf 78} (2008) 046007
[arXiv:0712.1396 [hep-th]];
\\
D.~L.~Jafferis and A.~Tomasiello,
``A simple class of N=3 gauge/gravity duals,''
JHEP {\bf 0810} (2008) 101
[arXiv:0808.0864 [hep-th]].

\bibitem{Koerber:2008rx}
 P.~Koerber, D.~Lust and D.~Tsimpis,
 JHEP {\bf 0807} (2008) 017
 [arXiv:0804.0614 [hep-th]].


\bibitem{Martelli:2008rt}
D.~Martelli and J.~Sparks,
``Notes on toric Sasaki-Einstein seven-manifolds and $AdS_4/CFT_3$,''
arXiv:0808.0904 [hep-th].



\bibitem{Duff:1987bx}
M.~J.~Duff, P.~S.~Howe, T.~Inami and K.~S.~Stelle,
``Superstrings in D = 10 from supermembranes in D = 11,''
Phys.\ Lett.\  B {\bf 191} (1987) 70.

\bibitem{Howe:2004ib}
P.~S.~Howe and E.~Sezgin,
``The supermembrane revisited,''
Class.\ Quant.\ Grav.\  {\bf 22} (2005) 2167
[arXiv:hep-th/0412245].


\bibitem{Carr:1986tk}
J.~L.~Carr, S.~J.~J.~Gates and R.~N.~Oerter,
``D = 10, N=2a Supergravity In Superspace,''
Phys.\ Lett.\  B {\bf 189} (1987) 68.


\bibitem{Gubser:2002tv}
S.~S.~Gubser, I.~R.~Klebanov and A.~M.~Polyakov,
``A semi-classical limit of the gauge/string correspondence,''
Nucl.\ Phys.\  B {\bf 636}, 99 (2002)
[arXiv:hep-th/0204051].
\bibitem{Frolov:2002av}
S.~Frolov and A.~A.~Tseytlin,
``Semiclassical quantization of rotating superstring in AdS(5) x S(5),''
JHEP {\bf 0206}, 007 (2002)
[arXiv:hep-th/0204226];
\\
S.~Frolov and A.~A.~Tseytlin,
``Multi-spin string solutions in AdS(5) x S**5,''
Nucl.\ Phys.\  B {\bf 668}, 77 (2003)
[arXiv:hep-th/0304255];
\\
S.~A.~Frolov, I.~Y.~Park and A.~A.~Tseytlin,
``On one-loop correction to energy of spinning strings in S(5),''
Phys.\ Rev.\  D {\bf 71}, 026006 (2005)
[arXiv:hep-th/0408187];
\\
S.~Frolov, A.~Tirziu and A.~A.~Tseytlin,
``Logarithmic corrections to higher twist scaling at strong coupling from
AdS/CFT,''
Nucl.\ Phys.\  B {\bf 766} (2007) 232
[arXiv:hep-th/0611269].

\bibitem{Mandal:2002fs}
G.~Mandal, N.~V.~Suryanarayana and S.~R.~Wadia,
``Aspects of semiclassical strings in AdS(5),''
Phys.\ Lett.\  B {\bf 543}, 81 (2002)
[arXiv:hep-th/0206103].

\bibitem{Gromov:2008bz}
N.~Gromov and P.~Vieira,
``The AdS4/CFT3 algebraic curve,''
arXiv:0807.0437 [hep-th].


\bibitem{Eichenherr:1981sk}
H.~Eichenherr and M.~Forger,
``Higher Local Conservation Laws For Nonlinear Sigma Models On Symmetric
Spaces,''
Commun.\ Math.\ Phys.\  {\bf 82}, 227 (1981).


\bibitem{Minahan:2008hf}
J.~A.~Minahan and K.~Zarembo,
``The Bethe ansatz for superconformal Chern-Simons,''
JHEP {\bf 0809}, 040 (2008)
[arXiv:0806.3951 [hep-th]];

\bibitem{Gaiotto:2008cg}
D.~Gaiotto, S.~Giombi and X.~Yin,
``Spin Chains in N=6 Superconformal Chern-Simons-Matter Theory,''
arXiv:0806.4589 [hep-th];

\bibitem{Grignani:2008is}
 G.~Grignani, T.~Harmark and M.~Orselli,
 ``The SU(2) x SU(2) sector in the string dual of N=6 superconformal
 Chern-Simons theory,''
 arXiv:0806.4959 [hep-th];
 \\
 G.~Grignani, T.~Harmark, M.~Orselli and G.~W.~Semenoff,
 ``Finite size Giant Magnons in the string dual of N=6 superconformal
 Chern-Simons theory,''
 arXiv:0807.0205 [hep-th];
 \\
 D.~Astolfi, V.~G.~M.~Puletti, G.~Grignani, T.~Harmark and M.~Orselli,
 ``Finite-size corrections in the SU(2) x SU(2) sector of type IIA string
 theory on AdS(4) x CP(3),''
 arXiv:0807.1527 [hep-th].


\bibitem{Bak:2008cp}
D.~Bak and S.~J.~Rey,
``Integrable Spin Chain in Superconformal Chern-Simons Theory,''
JHEP {\bf 0810}, 053 (2008)
[arXiv:0807.2063 [hep-th]].


\bibitem{Gromov:2008qe}
N.~Gromov and P.~Vieira,
``The all loop AdS4/CFT3 Bethe ansatz,''
arXiv:0807.0777 [hep-th].

\bibitem{Ahn:2008aa}
C.~Ahn and R.~I.~Nepomechie,
``N=6 super Chern-Simons theory S-matrix and all-loop Bethe ansatz
equations,''
JHEP {\bf 0809} (2008) 010
[arXiv:0807.1924 [hep-th]].



\bibitem{Mueller-Hoissen}
F. M\"uller--Hoissen and Richard St\"uckl, Coset spaces and
ten--dimensional unified theories. Class.\ Quant.\ Grav.\  {\bf 5}
(1988) 27.

\bibitem{Hatsuda:2002kx}
M.~Hatsuda, K.~Kamimura and M.~Sakaguchi,
``Super-PP-wave algebra from super-AdS x S algebras in eleven-dimensions,''
Nucl.\ Phys.\  B {\bf 637} (2002) 168
[arXiv:hep-th/0204002].


\end{thebibliography}
\end{document}